\documentclass[11pt, a4paper]{article}
\usepackage[top=1in, bottom=1in, left=1.25in, right=1.25in]{geometry}
\usepackage{booktabs}
\usepackage{multirow}
\usepackage{setspace}
\usepackage{natbib}
\usepackage{authblk}
\usepackage[pdftex]{graphicx}
\usepackage{amsmath}
\usepackage{amsthm}
\usepackage{mathtools}
\usepackage{amsfonts}
\usepackage{ascmac}
\usepackage{amssymb}
\usepackage{bm}
\usepackage{color}
\usepackage{enumitem}
\usepackage{mathrsfs}
\usepackage[ 
	colorlinks  = true, 
	anchorcolor = black,
	linkcolor   = blue,
	citecolor   = blue, 
	filecolor   = black, 
	urlcolor    = black
]{hyperref}
\usepackage{comment}
\usepackage{here}
\usepackage{enumitem}
\usepackage{longtable}
\usepackage{tabularx}
\usepackage{colortab}
\usepackage{colortbl}
\usepackage{arydshln}

\newcommand{\biblist}{\begin{list}{}
{\listparindent 0.0cm \leftmargin 0.50cm \itemindent -0.50 cm
\labelwidth 0 cm \labelsep 0.50 cm
\usecounter{list}}\clubpenalty4000\widowpenalty4000}
\newcommand{\ebiblist}{\end{list}}

\theoremstyle{plain}

\theoremstyle{definition}

\theoremstyle{remark}

\newcommand{\Var}{{\rm Var}}

\newcommand{\Bin}{{\rm Bin}}

\title{\bf Improving Maximum Tolerated Dose Selection in Model-Assisted Designs for Phase I Trials through Bayesian Dose-Response Model}
\author[1]{Rentaro Wakayama}
\author[2]{Tomotaka Momozaki}
\author[2]{Shuji Ando}
\affil[1]{Department of  Information Sciences, Graduate School of Science and Technology, Tokyo University of Science}
\affil[2]{Department of  Information Sciences, Faculty of Science and Technology, Tokyo University of Science}

\date{Last update: \today}

\begin{document}
\maketitle
\begin{abstract}
Model-assisted designs have garnered significant attention in recent years due to their high accuracy in identifying the maximum tolerated dose (MTD) and their operational simplicity. 
To identify the MTD, they employ estimated dose limiting toxicity (DLT) probabilities via isotonic regression with pool-adjacent violators algorithm (PAVA) after trials have been completed. 
PAVA adjusts independently estimated DLT probabilities with the Bayesian binomial model at each dose level using posterior variances ensure the monotonicity that toxicity increases with dose. 
However, in small sample settings such as Phase I oncology trials, this approach can lead to unstable DLT probability estimates and reduce MTD selection accuracy. 
To address this problem, we propose a novel MTD identification strategy in model-assisted designs that leverages a Bayesian dose-response model. 
Employing the dose-response model allows for stable estimation of the DLT probabilities under the monotonicity by borrowing information across dose levels, leading to an improvement in MTD identification accuracy. 
We discuss the specification of prior distributions that can incorporate information from similar trials or the absence of such information. 
We examine dose-response models employing logit, log-log, and complementary log-log link functions to assess the impact of link function differences on the accuracy of MTD selection. 
Through extensive simulations, we demonstrate that the proposed approach improves MTD selection accuracy by more than 10\% in some scenarios and by approximately 6\% on average compared to conventional approach. 
These findings indicate that the proposed approach can contribute to further enhancing the efficiency of Phase I oncology trials. 
\end{abstract}


\medskip

\section{Introduction} \label{sec:intro}
Since Phase I oncology trials are the first stage in which a drug candidate, previously confirmed to be safe in preclinical studies, is administered to humans (cancer patients), one of their key objectives is to identify the appropriate dosage, or recommended dose, for subsequent Phase II and Phase III trials. 
To achieve this, it is necessary to consider the relationship between drug efficacy and safety. 
The therapeutic efficacy of anticancer drugs has been observed to increase with escalating doses.
However, dose escalation also heighten the risk of severe adverse events (hereafter referred to as ``toxicity''). 
A monotonic relationship between dose and toxicity is generally assumed in oncology research.
Accordingly, the determination of the recommended dose relies on the probability of dose-limiting toxicity (DLT probability). 
The maximum tolerable toxicity probability, known as the target DLT probability, is conventionally set at around 0.3. 
This probability serves as the basis for identifying the maximum tolerated dose (MTD).
In Phase I oncology trials, multiple dose levels are predefined during the study design phase, and the MTD is identified using established dose-finding methodologies. 
For a comprehensive review of dose-finding methods in Phase I oncology trials, see \cite{chevret2006}, \cite{ting2006dose}, and \cite{gerke2008optimal}.

Algorithm-based designs, such as the 3+3 design \citep{storer1989design}, are still used as the mainstream dose-finding method due to their simplicity \citep{araujo2021contemporary}.
Algorithm-based designs determine dose escalation and de-escalation during the trial using simple, predefined algorithms rather than assuming a model for dose and toxicity.
Therefore, the implementation of clinical trials with Algorithm-based designs does not require complex computer calculations.
However, the 3+3 design has the drawback of frequently stopping at doses lower than the MTD and determining the MTD based solely on the data from six patients treated at that dose \citep{o1991methods,le2009dose}. 
Furthermore, the prioritization of simplicity in the 3+3 design results in several practical limitations, such as the lack of flexibility to reevaluate sample sizes \citep{liu2020i3+,paoletti2015statistical}. 
For similar reasons, extensions of the 3+3 design, such as the rolling six design \citep{skolnik2008shortening}, also suffer from low accuracy in identifying the MTD \citep{onar2010simulation,zhao2011superiority}. 

To address the limitations of algorithm-based designs, particularly their low accuracy in identifying the MTD, model-based designs, exemplified by the continuous reassessment method \citep[CRM;][]{o1990continual}, have been proposed as a more statistically rigorous alternative.
In contrast to algorithm-based designs, model-based designs determine dose escalation and de-escalation using a predefined dose-toxicity model (e.g., the power model and the logit model).
Utilizing information from all dose levels and iteratively updating the estimated dose-toxicity relationship, patients can be administered the dose that is most likely to represent the optimal MTD at that time. 
Although this sequential approach improves the accuracy of MTD identification, even in Phase I oncology trials with a small patient cohort, compared to algorithm-based methods \citep{green2010clinical}, the statistical and computational complexity arising from iterative model updates makes dose allocation appear as a ``black box'' to practitioners, leading to the highly restricted use of model-based designs \citep{araujo2021contemporary}. 

Given these limitations, model-assisted designs \citep{zhou2018comparative, yuan2019model}, which merge the operational simplicity of algorithm-based designs with the good performance of model-based approaches, have attracted growing interest in recent years. 
Model-assisted designs identify the MTD by estimating DLT probabilities through the isotonic regression with post-trial data on the number of patients treated and the number of observed DLT events at each dose level. 
The pool-adjacent violators algorithm (PAVA) \citep{barlow1972statistical, leeuw2009isotone} is utilized to estimate the DLT probabilities with weighted means based on the variance of the estimates, ensuring compliance with the natural monotonicity assumption in Phase I oncology trials that toxicity increases with dose. 
Since the isotonic regression independently estimates the DLT probability for each dose level, estimates tend to be unstable in Phase I oncology trials due to the typically small sample sizes.
Additionally, the variances of the estimates used as weights for monotonicity in PAVA are similarly affected. 

To overcome the limitations of DLT probability estimation using the isotonic regression with PAVA, we propose a novel MTD identification strategy in model-assisted designs that leverages a Bayesian dose-response model, specifically a binomial regression with the number of observed DLTs as the response variable and the dose level as the explanatory variable. 
By using logit, log-log, or complementary log-log (clog-log) link functions, DLT probability estimation naturally accounts for the monotonicity between dose and toxicity. This contrasts with PAVA, which requires ad hoc adjustments to the estimates.
Additionally, the dose-response model allows for the estimation of DLT probabilities at dose levels with small sample size using information from dose levels with larger allocations, thereby enhancing the stability of MTD identification relative to conventional approaches. 
An additional advantage is that the estimation process considers not only the number of observed DLTs and patients allocated to each dose level but also the dose levels themselves. 
It is important to note that our approach improves the estimation of DLT probabilities in model-assisted designs using post-trial toxicity data to enhance MTD identification, while preserving the simplicity of trial implementation, unlike model-based designs. 

In this study, we focus on the Bayesian optimal interval (BOIN) design \citep{liu2015bayesian}, one of the most prominent model-assisted designs.
BOIN is also one of the interval designs \citep{ivanova2007cumulative, ivanova2009dose, ji2010modified}, where dose escalation and de-escalation are determined based on the observed DLT rate at the current dose, compared to a predefined toxicity tolerance interval to decide the appropriate dose for the next patient cohort. 
\cite{oron2011dose} demonstrate that interval designs offer convergence properties similar to those of the CRM. 
We conduct extensive simulations to examine the impact of using the dose-response model instead of the isotonic regression on MTD selection. 
Our emphasis on the BOIN design stems from its superior accuracy in identifying the optimal MTD and its lower risk of overdosing compared to the mTPI design, as well as its simpler implementation and ease of execution based on estimated DLT probabilities compared to the Keyboard design \citep{zhou2018comparative}. 
The BOIN framework has been extensively developed to address various clinical scenarios, with iBOIN \citep{zhou2021incorporating} (BOIN with informative prior) allowing incorporation of historical toxicity information through informative priors, gBOIN \citep{mu2019gboin} (generalized BOIN) generalizing the design to handle continuous, quasi-binary, and binary toxicity endpoints, U-BOIN \citep{zhou2019utility} (utility-based BOIN) incorporating utility functions for dose optimization based on risk-benefit trade-offs, and BOIN12 \citep{lin2020boin12} (BOIN phase I/II) providing one-stage utility-based design for targeted therapies and immunotherapies (see \cite{ying2023model-assisted} for comprehensive reviews of these BOIN variants). 
These developments demonstrate the adaptability of the model-assisted approach. 
However, regardless of these extensions, the core MTD selection process in BOIN designs relies on isotonic regression for final dose recommendation. 
Our proposed modification addresses this fundamental aspect by incorporating parametric dose-response modeling to potentially improve MTD selection accuracy across all BOIN variants. 

In Bayesian dose-response models, the choice of link function and the specification of prior distribution are critical factors influencing MTD selection. 
Compared to the logit link, the log-log link assumes lower DLT probabilities at lower dose levels, whereas the clog-log link assumes higher DLT probabilities at higher dose levels, suggesting that the appropriate link function should be chosen based on fundamental trial insights. 
Additionally, It may be preferable to incorporate information from similar trials into the prior distribution in small sample trials, whereas a non-informative prior is used in the absence of such information. 
Therefore, we also demonstrate the effects of prior distribution specification and different link function choices on MTD selection within our proposed method. 

The rest of this paper is organized as follows. 
Section \ref{sec:boin} provides a brief overview of the BOIN design and the method for estimating DLT probabilities using PAVA. 
Section \ref{sec:prop} propose the MTD identification strategy in BOIN using the Bayesian dose-response model and the specification of the prior distribution. 
Section \ref{sec:sim} demonstrates the performance of our approach compared to the conventional one.
Section \ref{sec:dis} discusses the dose-response model and the prior distribution specification based on insights obtained from the simulations. 
Section \ref{sec:concl} provides the conclusion and some remarks.

\section{Bayesian Optimal Interval design} \label{sec:boin}
The BOIN design \citep{liu2015bayesian} is one of the interval designs, where dose escalation and de-escalation are determined based on a simple comparison of the observed DLT rate at the current dose level with a predefined toxicity tolerance interval \citep{oron2011dose}. 
One of the key advantages of the BOIN design is that they do not require complex calculations to determine the dose for the next cohort during the trial. 
The BOIN design consists of three key steps. 
In the following, we provide a brief explanation of each step. 

\subsection{Preparation for trial} 
The BOIN design achieves optimization by obtaining the toxicity tolerance interval minimizing the probability of incorrect dose escalation and de-escalation decisions in the Bayesian framework. 
\cite{liu2015bayesian} propose the local BOIN design, optimized based on point hypotheses, and the global BOIN design, optimized based on interval hypotheses. 
Since they recommend the use of the local BOIN design due to its better performance, we also focus on the local BOIN design, hereafter simply referred to as the BOIN design. 

Letting $p_j$ denote the true DLT probability of dose level $j$ ($j = 1,\dots,J$), the BOIN design defines three point hypotheses 
\begin{equation*}
    H_{0j}: p_j = \phi, \quad H_{1j}: p_j = \phi_1, \quad H_{2j}: p_j = \phi_2, 
\end{equation*}
where $\phi_1$ is the highest DLT probability for dose escalation, indicating substantially underdosing (i.e., below the MTD), and $\phi_2$ is the lowest DLT probability for dose de-escalation, indicating substantially overdosing (i.e., above the MTD). 
$H_{0j}$ indicates that the current dose is the MTD, suggesting that the same dose should be retained for the next cohort. 
$H_{1j}$ indicates that the current dose is subtherapeutic (below the MTD), necessitating an escalation in dose. 
$H_{2j}$ indicates that the current dose is excessive (above the MTD), requiring a dose de-escalation. 
Table \ref{tb:hypo} summarizes the three point hypotheses for the current dose retention, dose escalation, and dose de-escalation using $\mathcal{R}, \mathcal{E}$, and $\mathcal{D}$, respectively, where $\bar{\mathcal{R}}, \bar{\mathcal{E}}$, and $\bar{\mathcal{D}}$ represent events that make incorrect decisions under each point hypotheses.

\begin{table}[H]
    \centering
    \caption{Three hypotheses in the BOIN design and their corresponding correct and incorrect dose selections.}
    \vspace{2mm}
    \label{tb:hypo}
    \scalebox{0.95}{
    \begin{tabular}{ccc}
    \hline
    $H_{0j}: p_j=\phi$ & $H_{1j}: p_j=\phi_1$ & $H_{2j}: p_j=\phi_2$ \\ \hline
    $\mathcal{R}:$ retainment & $\mathcal{E}:$ escalation & $\mathcal{D}:$ de-escalation \\
    $\bar{\mathcal{R}}:$ escalation, de-escalation & $\bar{\mathcal{E}}:$ retainment, de-escalation & $\bar{\mathcal{D}}:$ retainment, escalation \\ \hline
    \end{tabular}
    }
\end{table}

Under the Bayesian framework, letting $\pi_{ij}=P(H_{ij})$ $(i=0,1,2)$ denote the prior probability that each hypothesis is true, the probability of making an incorrect decision at each of the dose assignments is given by 
\begin{align*}
    P(\phi-\Delta_L,\;\phi+\Delta_U) 
    =& \pi_{0j} \Bin(n_j(\phi-\Delta_L);n_j,\phi)-\pi_{1j} \Bin(n_j(\phi-\Delta_L);n_j,\phi_1) \\
    &+ \pi_{2j} \Bin(n_j(\phi+\Delta_U);n_j,\phi_2)-\pi_{0j} \Bin(n_j(\phi+\Delta_U);n_j,\phi) \\
    &+\pi_{0j}+\pi_{1j} \\
    =& P(\phi-\Delta_L)+P(\phi+\Delta_U)+\pi_{0j}+\pi_{1j},
\end{align*}
where $\phi+\Delta_U$ and $\phi-\Delta_L$ are the upper and lower bounds of the toxicity tolerance interval, respectively, $n_j$ is the number of the allocated patients at the dose level $j$, $\Bin(x;n,p)$ is the cumulative distribution function of the binomial distribution with size and probability parameters $n$ and $p$ evaluated at the value $x$, and 
\begin{align*}
P(\phi-\Delta_L) &= \pi_{0j} \Bin(n_j(\phi-\Delta_L);n_j,\phi)-\pi_{1j} \Bin(n_j(\phi-\Delta_L);n_j,\phi_1), \\
P(\phi+\Delta_U) &= \pi_{2j} \Bin(n_j(\phi+\Delta_U);n_j,\phi_2)-\pi_{0j} \Bin(n_j(\phi+\Delta_U);n_j,\phi). 
\end{align*}
Hence, by minimizing $P(\phi-\Delta_L,\phi+\Delta_U)$, that is, by minimizing $P(\phi-\Delta_L)$ and $P(\phi+\Delta_U)$, the upper and lower bounds of the toxicity tolerance interval are derived as
\begin{equation*}
\phi-\Delta_L = \frac{\log\left(\frac{1-\phi_1}{1-\phi}\right)}{\log\left(\frac{\phi(1-\phi_1)}{\phi_1(1-\phi)}\right)}, \qquad
\phi+\Delta_U = \frac{\log\left(\frac{1-\phi}{1-\phi_2}\right)}{\log \left(\frac{\phi_2(1-\phi)}{\phi(1-\phi_2)}\right)} 
\end{equation*}
with the non-informative prior $\pi_{0j}=\pi_{1j}=\pi_{2j}=1/3$. 
For detailed derivations, see the original BOIN paper or the author's book \citep{liu2015bayesian, ying2023model-assisted}. 
These bounds are independent of the dose level $j$ and the number of subjects $n_j$. 
The specification of $\phi_1$ and $\phi_2$ has been extensively examined and it is recommended to set $(\phi_1,\phi_2)=(0.6\phi, 1.4\phi)$ \citep{liu2015bayesian, ying2023model-assisted}. 
Therefore, the optimal toxicity tolerance interval for dose escalation and de-escalation can be prespecified during trial planning only by setting the target DLT probability $\phi$.

\subsection{Collected toxicity data}
In the BOIN design, the trial proceeds using the toxicity tolerance interval described in the previous section, while data on toxicity in patients are gathered. 
Table \ref{tb:data} shows a dose escalation and de-escalation rule for the BOIN design with $\phi=0.3$, considering a total of 36 patients divided into 12 cohorts, each consisting of 3 patients. 
At a current dose level, if the number of DLTs reaches the lower number, the dose is escalated in the next cohort, whereas if it reaches the upper number, the dose is de-escalated. 
Otherwise, the current dose is retained.
Upon completing the trial according to this rule, the DLT probability is estimated and the MTD is selected based on the total observed numbers of DLTs and allocated patients at each dose level. 
Note that dose levels that have not been administered to any patients and those where toxicity data meet the exclusion criteria after completing treatment in the final cohort are excluded from consideration as MTD candidates. 

\begin{table}[h]
\centering
\caption{A dose escalation and de-escalation rule for the BOIN design with $\phi=0.3$ and a total of 36 patients, consisting of 12 cohorts with 3 patients each.}
\label{tb:data}
\vspace{2mm}
\begin{tabular}{c c c c c c c c}
\hline
Numbers of cumulative allocated patients & 3 & 6 & 9 & 12 & $\cdots$ & 33 & 36 \\ \hline 
Lower numbers of DLTs $(\phi-\Delta_L)$ & 0 & 1 & 2 & 2 & $\cdots$ & 7 & 8 \\
Upper numbers of DLTs $(\phi+\Delta_U)$ & 2 & 3 & 4 & 5 & $\cdots$ & 12 & 13 \\ \hline
\end{tabular}
\end{table}

\subsection{Estimation of DLT probability and MTD selection} \label{sec:pava} 
Let $m_j$ denote the total number of DLTs and allocated patients at dose $d_j$. 
In the BOIN design, the DLT probability is estimated using the isotonic regression with the PAVA \citep{barlow1972statistical, leeuw2009isotone}. 
The procedure is as follows. 
First, for all dose levels, we calculate
\begin{equation} \label{eq:hat_y}
    \hat{y}_j = \frac{m_j+0.05}{n_j+0.1}.
\end{equation} 
This represents the posterior mean of the DLT probability using a prior distribution of ${\rm Beta}(0.05, 0.05)$. 
In this process, the posterior probability that the DLT probability at a given dose level exceeds the target DLT probability is calculated. 
If this probability exceeds the prespecified threshold, the corresponding dose level is also excluded from the MTD candidates. 

When $\{\hat{y}_j\}$ violate the monotonicity condition, that is, $\hat{y}_{j-1} \geq \hat{y}_{j}$, they are adjusted to 
\begin{equation} \label{eq:f_estimation}
  \hat{p}_j = \hat{z}_j + j\times 1.0 \times 10^{-10}, 
\end{equation}
where $\hat{z}_j$ is the weighted mean 
\begin{equation} \label{eq:pooling}
    \hat{z}_j = \hat{z}_{j-1} = \frac{\hat{y}_j\frac{1}{\Var(\hat{y}_j)}+\hat{y}_{j-1}\frac{1}{\Var(\hat{y}_{j-1})}}{\frac{1}{\Var(\hat{y}_j)}+\frac{1}{\Var(\hat{y}_{j-1})}} 
\end{equation}
based on the variance of $\hat{y}_j$ 
\begin{equation*}
    \Var(\hat{y}_j) = \frac{(m_j+0.05)(n_j-m_j+0.05)}{(n_j+0.1)^2(n_j+0.1+1)}.
\end{equation*}
When $\{\hat{y}_j\}$ satisfies the monotonicity, the procedure in Equation \eqref{eq:pooling} can be skipped, and the estimated DLT probability $\hat{p}_j$ can be obtained by substituting the value of $\hat{y}_j$ into the value of $\hat{z}_j$ in equation \eqref{eq:f_estimation}. 
Finally, the dose level having the estimated DLT probability closest to $\phi$ is selected as the MTD. 

PAVA is commonly employed due to its ability to identify the optimal MTD while its simple and straightforward procedure. 
However, its performance remains somewhat questionable in small-sample settings, such as Phase I oncology trials. 
As shown in Equation \eqref{eq:hat_y}, the estimation of DLT probability is conducted independently for each dose level. 
As a result, dose levels with small $n_j$ can make the estimation unstable. 
Likewise, the variance of the estimates used to ensure monotonicity may also lead to unstable estimations. 
Given this context, we expect that PAVA may not be the optimal method for estimating DLT probabilities in Phase I oncology trials.

\section{Bayesian dose-response model for estimating DLT probabilities} \label{sec:prop} 
In this section, we propose an MTD selection method using a Bayesian dose-response model for DLT probability estimation to resolve the issues of using PAVA in the BOIN design under small-sample settings, such as those in Phase I oncology trials, discussed in the previous section. 
The dose-response model leverages all available data, the total number of DLTs $\{m_j\}$ and the allocated patients $\{n_j\}$ and dose $\{d_j\}$, allowing for the stable estimation of DLT probabilities at dose levels with smaller sample sizes by borrowing information from dose levels with larger sample sizes. 
This stability facilitates the accurate selection of the optimal MTD. 
This approach naturally incorporates monotonicity, where toxicity increases with dose, without requiring the ad hoc adjustments in PAVA. 
Section \ref{sec:dr_model} describes the dose-response model used in this study. 
Section \ref{sec:prior} discusses the specification of prior distributions for this model. 
Note that, unlike model-based designs, our approach does not use the dose-response model for making dose escalation and de-escalation decisions, but for estimating the DLT probabilities. 
Specifically, as in \cite{liu2015bayesian}, the dose level whose estimated DLT probability based on the dose-response model is closest to the target DLT probability $\phi$ is selected as the MTD. 
Thus, it facilitates more precise MTD selection than conventional approaches while retaining the practical advantages of the BOIN design.

\subsection{Dose-response model} \label{sec:dr_model} 
Assuming $m_j \sim \Bin(n_j, \pi(d_j))$, we consider the dose-response model, used in \cite{neuenschwander2008critical}, 
\begin{equation} \label{eq:drmodel}
    g(\pi(d_j)) = \beta_0 + \exp(\beta_1)\log\left(\frac{d_j}{d^*}\right), 
\end{equation}
where $g(\cdot)$ is the link function, $\beta_0$ and $\beta_1$ are coefficient parameters, and $\pi(d_j)$ is the DLT probability for dose $d_j$. 
The $d^*$, referred to as the reference dose, is selected from $\{d_j\}$ (e.g., $d^*=d_3$). 
With the logit link function, $\beta_0$ in the dose-response model \eqref{eq:drmodel} is equal to the log-odds of the DLT probability at dose $d^*$. 
Additionally, the log odds ratio of $\pi(d_i)$ and $\pi(d_j)$ is proportional to the exponential of the coefficient parameter $\beta_1$, that is, 
\begin{equation*}
    \exp(\beta_1) = \frac{\mathrm{logit}\{ \pi(d_j) \} - \mathrm{logit}\{ \pi(d_i) \}}{\log(d_j/d_i)}.
\end{equation*}
The reason for using the dose-response model \eqref{eq:drmodel} with the reference dose is not only to improve parameter interpretability, as mentioned earlier, but also to reduce the estimation variability due to differences in dose scaling. 
Moreover, as described later, it facilitates the specification of prior distributions. 

\begin{figure}[H]
    \centering
    \includegraphics[width=\columnwidth]{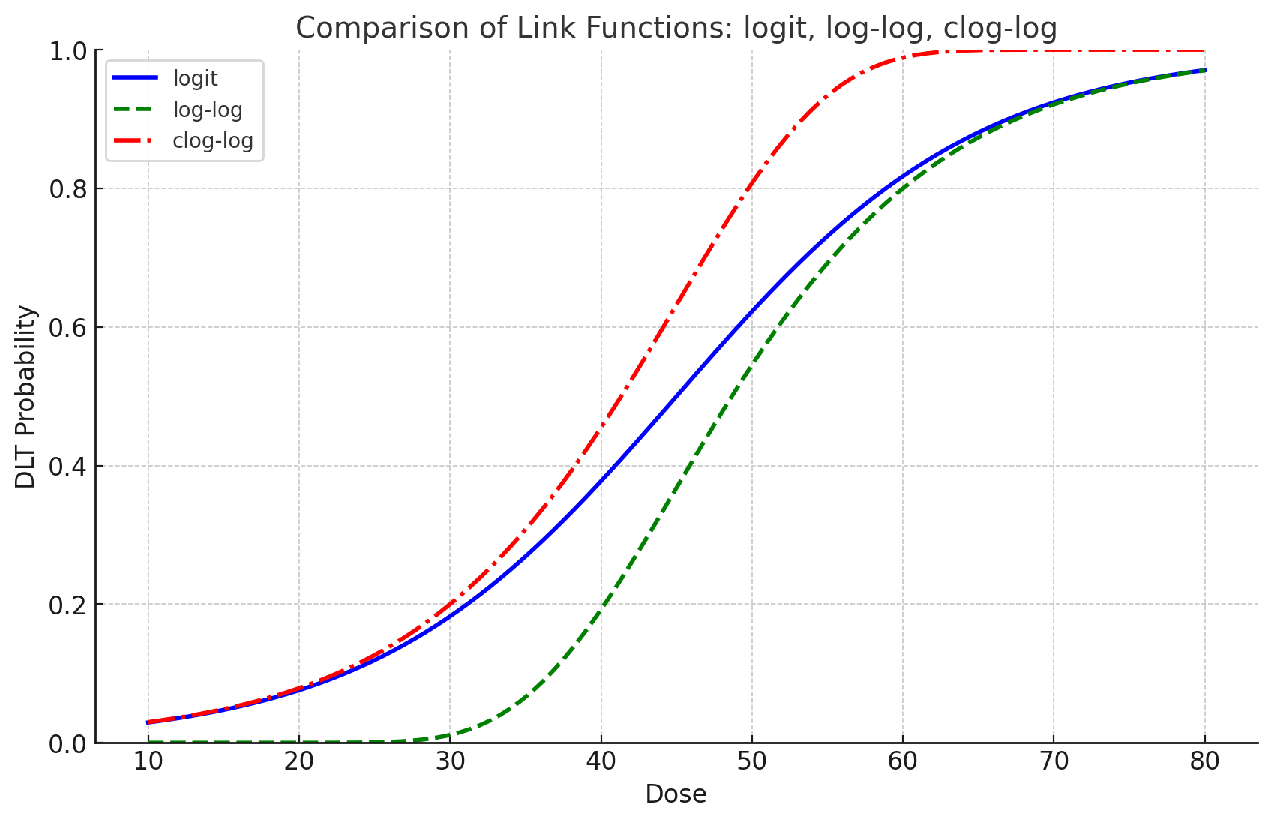}
    \caption{The dose-response curves representing DLT probabilities described by the three link functions: logit, log-log, and clog-log.}
    \label{fig:link}
\end{figure}

In addition to the logit link, this study explores the use of log-log and clog-log link functions described as   
\begin{equation*}
    g(\pi(d_j)) = -\log\{ -\log(\pi(d_j)) \} ~~ \mbox{and} ~~ g(\pi(d_j)) = \log\{ -\log(1-\pi(d_j)) \}, 
\end{equation*}
respectively. 
Figure \ref{fig:link} visually illustrates the differences among the three link functions. 
Compared to the logit link, the dose-response model using the log-log link demonstrates a more rapid decrease in DLT probability as the dose decreases, making it more likely for the DLT probability to approach zero at lower doses. 
On the other hand, the clog-log link leads to a more rapid increase in DLT probability at higher doses, making it more likely to approach one. 
Recognizing the differences in dose-response curve characteristics induced by various link functions, we assess the impact of these link functions on MTD selection accuracy. 

\subsection{Prior distribution for dose-response model} \label{sec:prior} 
The specification of prior distributions is crucial for incorporating available information about the dose-toxicity relationship, or reflecting the absence of such information. 
Particularly in cases with limited sample sizes, leveraging reliable external information can enhance estimation accuracy. 
In the absence of prior information, flat prior distributions or those with large variance are typically used. 
Assigning such priors to the parameters in the dose-response model \eqref{eq:drmodel} leads to a U-shaped prior for DLT probabilities, with spikes at zero and one. 
While this may be appropriate for the highest and lowest doses, it may result in a misspecified prior for intermediate dose levels, where moderate toxicity rates are typically expected rather than extreme values near zero or one. 
This prior-data mismatch can lead to biased estimation, which is particularly problematic given the limited sample sizes characteristic of phase I trials. 
To avoid this issue, following \cite{neuenschwander2008critical}, we consider approximating the prior distribution for DLT probabilities obtained through the dose-response model \eqref{eq:drmodel} with dose-specific prior distributions for DLT probabilities at each dose level. 
Specifically, we first set quantile targets for DLT probabilities at each dose based on clinical knowledge or design constraints, then derive the hyperparameters of the coefficient priors by minimizing the discrepancy between these target quantiles and the quantiles implied by the dose-response model. 

Let $q_{d_j} = \{ q_{d_j}(p_1),\dots,q_{d_j}(p_K) \}$ denote the set of $K$ quantiles for the dose-specific prior distribution of the DLT probability $\pi(d_j)$ at dose level $j$, such that $P\{\pi(d_j) \leq q_{d_j}(p_k)\} = p_k$ for $k = 1, \ldots, K$. 
Our prior information is thus summarized by $J \times K$ quantiles, $Q = (q_{jk})$ where $q_{jk} = q_{d_j}(p_k)$ for $j = 1, \ldots, J$ and $k = 1, \ldots, K$. 
On the other hand, let $Q' = (q'_{jk})$ represent the corresponding quantiles numerically computed from the dose-response model \eqref{eq:drmodel} using the prior distributions $\beta_0 \sim N(\gamma_0, \sigma_0^2)$ and $\beta_1 \sim N(\gamma_1, \sigma_1^2)$. 
That is, each element $q'_{jk}$ represents the $k$-th quantile of $\pi(d_j) = g^{-1}\{\beta_0 + \exp(\beta_1) \log(d_j/d^*)\}$ under the assumed priors for coefficients. 
Our aim is to determine hyperparameters $\eta = \{\gamma_0, \gamma_1, \sigma_0^2, \sigma_1^2\}$ of the coefficient priors that most faithfully reproduce the dose-specific prior information. 
To achieve this, we minimize the discrepancy $C(Q, Q')$ between the target quantiles $Q$ and the model-implied quantiles $Q'$ by solving the optimization problem $\eta^* = {\rm argmin}_{\eta} C(Q, Q')$. 
We define the loss function as 
\begin{equation} \label{eq:obje}
    C(Q, Q') = \sum_{j,k}(q_{jk}-q'_{jk})^2. 
\end{equation}
While \cite{neuenschwander2008critical} employ $C(Q, Q') = \max_{j,k} |q_{jk} - q'_{jk}|$, which focuses on minimizing the largest quantile error, it suffers from convergence issues caused by minimizing the maximum absolute value using stochastic optimization based on the Metropolis algorithm \citep{robert2004monte}. 
In contrast, our squared error approach considers deviations across all quantiles simultaneously rather than focusing solely on the worst-case error. 
The differentiability of the loss function \eqref{eq:obje} enhances optimization efficiency while allowing the evaluation of errors across all quantiles, leading us to select it as our preferred loss function. 

In cases where specific prior information about quantiles $Q$ is unavailable, we set the target quantiles $Q$ based on minimally informative unimodal Beta distributions defined by \cite{neuenschwander2008critical} 
The minimally informative unimodal Beta distribution is defined as the Beta distribution $\pi \sim \text{Beta}(a,b)$ that satisfies (i) $P\{\pi < q(p)\} = p$ with the least information ((ii) minimal $a + b$) among all unimodal Beta distributions (where (iii) $a \geq 1$ or $b \geq 1$). 
A key advantage of this prior distribution is that it requires only a single quantile criterion for each dose level about condition (i) to fully specify the prior distribution. 
When $q(p) > p$, corresponding to a right-skewed prior, the hyperparameter $a$ of the minimally informative unimodal Beta distribution $\text{Beta}(a,1)$ is given by $a = \log(p)/\log\{q(p)\}$. 
Conversely, when $q(p) < p$, indicating a left-skewed prior, the hyperparameter $b$ of the minimally informative unimodal Beta distribution $\text{Beta}(1,b)$ is given by $b = \log(1-p)/\log\{1-q(p)\}$. 

By leveraging the dose-response model \eqref{eq:drmodel}, we can further reduce the information required for prior specification. 
In the following steps, we proceed to obtain the prior quantiles $Q$ for the DLT probability using the minimally informative unimodal Beta distribution and specify the hyperparameters $\eta$ of the coefficient priors to match these target quantiles through the optimization procedure described earlier. 
\begin{enumerate}[label=Step \arabic*., leftmargin=*]
    \item 
    Set $q_1$ and $q_J$ such that $P(\pi(d_1)>q_1) = p_1$ and $P(\pi(d_J) \leq q_J) = p_J$ for the lowest and highest doses $d_1$ and $d_J$, respectively. 
    \item 
    Compute the prior medians $\mu_1$ and $\mu_J$ for the minimally informative unimodal Beta distribution at doses $d_1$ and $d_J$ using $q_1$ and $q_J$ from Step 1. 
    Specifically, when $q_1>p_1$, calculate $a_1=\log(p_1)/\log(q_1)$ and obtain $\mu_1 = 2^{-1/a_1}$, whereas when $q_1<p_1$, calculate $b_1=\log(1-p_1)/\log(1-q_1)$ and obtain $\mu_1 = 1-2^{-1/b_1}$. 
    The same procedure is followed for dose $d_J$. 
    \item 
    Solve for $\beta'_0$ and $\beta'_1$ that satisfy $g(\mu_1) = \beta_0 + \exp(\beta_1) \log(d_1/d^*)$ and $g(\mu_J) = \beta_0 + \exp(\beta_1) \log(d_J/d^*)$ based on the dose-response model \eqref{eq:drmodel}. 
    \item 
    Compute the prior median $\mu_j = g^{-1}\{\beta'_0 + \exp(\beta'_1) \log(d_j/d^*)\}$ for $j=2,\ldots,J-1$ and derive the hyperparameter of the minimally informative unimodal Beta prior. 
    \item 
    Calculate the quantiles $Q$ of the minimally informative unimodal Beta prior at each dose level using the results from Steps 1 through 4, and optimize the hyperparameter $\eta$ that minimizes the loss function \eqref{eq:obje}. 
\end{enumerate}
In Step 2, the complete prior specification across all dose levels can be achieved by considering BOIN's safety mechanisms and specifying only two probabilities, $p_1 = P\{\pi(d_1) > \phi(=0.3)\}$ for the lowest dose and $p_J = P\{\pi(d_J) \leq 0.3\}$ for the highest dose. 
This approach significantly reduces the burden of prior specification, eliminating the need for detailed prior information for each dose individually. 
The BOIN design imposes a dose elimination/safety stopping rule. 
According to this rule, the dose level $j$ and the higher doses are eliminated from the trial if the posterior probability $P\{\pi(d_j) > 0.3 | m_j, n_j\} > 0.95$ and $n_j \geq 3$ \citep{yan2020boin, ying2023model-assisted}. 
The trial is stopped if the lowest dose is eliminated. 
This safety mechanism provides natural constraints for our prior specification, $p_1 \leq 0.95$ and $1 - p_J \leq 0.95$, that is, $p_J \geq 0.05$. 
Within these constraints, we address the dual concerns of controlling unacceptable toxicity risk at the lowest dose and avoiding underdosing at the highest dose \citep{neuenschwander2008critical}. 
For the lowest dose $d_1$, we set $p_1 = P\{\pi(d_1) > 0.3\} = 0.05$, prioritizing control of unacceptable toxicity occurrence. 
For the highest dose $d_J$, we set $p_J = P\{\pi(d_J) \leq 0.3\} = 0.21$ yielding $P\{\pi(d_J) \leq 0.1\} = 0.05$, thereby avoiding underdosing. 
While these values align with BOIN's safety framework, practitioners should tailor these probability specifications to each trial's scientific and clinical context, considering regulatory guidance and ethical committee requirements. 
To examine the impact of different probability choices on MTD selection performance, sensitivity analyses with alternative $p_1$ and $p_J$ values are provided in Appendix.

\section{Simulation studies} \label{sec:sim} 
This section evaluates the accuracy of MTD selection in the proposed method compared to conventional approaches through comprehensive simulations under various scenarios. 
Section \ref{sec:setting} provides details on the simulation settings. 
Section \ref{sec:sim_prior} discusses the prior distributions used in the simulations based on Section \ref{sec:prior} and describes the posterior computation algorithm implemented. 
Section \ref{sec:result} presents the simulation results.

\subsection{Simulation setting} \label{sec:setting}
We assume a trial design with 3 patients per cohort, 12 cohorts in total, and 36 participants overall. 
The dose levels used in the trial are a six-level dose set based on a modified Fibonacci sequence: $(10, 20, 30, 45, 60, 80)$ mg. 
The target DLT probability, $\phi$, is set to $\phi = 0.3$.
Table \ref{tab:scenarios} presents the eight DLT probability scenarios used in our simulations. 
Scenarios 1 through 4 are intentionally designed to be unfavorable for the dose-response model, with steep or shallow gradients. 
In contrast, scenarios 5 through 8 are obtained based on the dose-response model \eqref{eq:drmodel} with $d^* = d_3$ (scenarios 5 and 6 for the logit model, scenario 7 for the log-log model, and scenario 8 for the complementary log-log model). 
Under these settings, we generate 1000 datasets for each scenario and calculate the proportion of dose levels selected as the MTD by each method. 

\begin{table}[H]
    \centering
    \caption{Dose levels and DLT probabilities in simulation scenarios. 
    The scenarios from 5 to 8 are set based on the dose-response model \eqref{eq:drmodel}. 
    The optimal MTD is highlighted in bold.}
    \label{tab:scenarios}
    \vspace{2mm}
    \begin{tabular}{ccccccc} \hline
           Dose levels & 1 & 2 & 3 & 4 & 5 & 6  \\ 
           Doses & 10 & 20 & 30 & 45 & 60 & 80  \\ \hline 
           scenario 1 & 0.02 & 0.15 & 0.20 & $\bm{0.30}$ & 0.35 & 0.55 \\
           scenario 2 & 0.01 & 0.04 & 0.14 & 0.18 & 0.22 & $\bm{0.30}$ \\
           scenario 3 & 0.01 & 0.03 & 0.10 & 0.20 & $\bm{0.30}$ & 0.55 \\
           scenario 4 & 0.15 & $\bm{0.30}$ & 0.36 & 0.50 & 0.55 & 0.64 \\ \hdashline
           scenario 5 & 0.08 & 0.19 & $\bm{0.30}$ & 0.44 & 0.54 & 0.64 \\
           scenario 6 & 0.03 & 0.09 & 0.17 & $\bm{0.30}$ & 0.42 & 0.55 \\
           scenario 7 & 0.09 & $\bm{0.30}$ & 0.45 & 0.59 & 0.68 & 0.75 \\
           scenario 8 & 0.08 & 0.19 & $\bm{0.30}$ & 0.46 & 0.60 & 0.75 \\ \hline
     \end{tabular}
\end{table}

\subsection{Prior distribution of parameters and posterior computation} \label{sec:sim_prior}
The prior distributions for the regression parameters $\beta_0$ and $\beta_1$ are specified following the approach presented in Section \ref{sec:prior}. 
In our simulation study, we set the number of quantiles used at each dose level to $K=3$, selecting the 2.5\%, 50\%, and 97.5\% quantiles. 
To ensure that simulations are conducted under equal conditions, we adopt the specification of the prior distribution based on the minimally informative unimodal Beta prior. 
Through the procedure in Section \ref{sec:prior}, we obtain the quantiles for the logit, log-log, and clog-log links are 
\begin{align*}  
    q_{j1} &= \{0.00, 0.01, 0.01, 0.02, 0.03, 0.06\}, \\  
    q_{j2} &= \{0.08, 0.18, 0.27, 0.40, 0.49, 0.59\}, \\
    q_{j3} &= \{0.36, 0.65, 0.82, 0.93, 0.97, 0.98\},  
\end{align*}  
\begin{align*}  
    q_{j1} &= \{0.00,0.01,0.01,0.02,0.03,0.06\}, \\  
    q_{j2} &= \{0.08,0.22,0.33,0.44,0.52,0.59\}, \\
    q_{j3} &= \{0.36,0.74,0.88,0.95,0.98,0.98\},  
\end{align*}  
and
\begin{align*}  
    q_{j1} &= \{0.00,0.01,0.01,0.02,0.02,0.06\}, \\  
    q_{j2} &= \{0.08,0.17,0.25,0.37,0.47,0.59\}, \\
    q_{j3} &= \{0.36,0.62,0.79,0.91,0.97,0.98\},  
\end{align*}  
respectively. 
Consequently, the prior distributions for the coefficient parameters $\beta_0$ and $\beta_1$ in the logit, log-log, and clog-log link models are 
\begin{equation*}
    \beta_0 \sim N(-1.592,1.371), \quad \beta_1 \sim N(0.412,0.784), 
\end{equation*}
\begin{equation*}
    \beta_0 \sim N(-0.231,0.847), \quad \beta_1 \sim N(0.068,0.544), 
\end{equation*}
and 
\begin{equation*}
    \beta_0 \sim N(-1.549,0.943), \quad \beta_1 \sim N(0.142,0.743), 
\end{equation*}
respectively. 

Using the obtained prior, we perform Bayesian estimation of the DLT probability with the \texttt{cmdstanr} package \citep{cmdstanr2024}, which allows for executing the probabilistic programming language \texttt{Stan} \citep{carpenter2017stan} in \texttt{R} \citep{r2024}. 
The \texttt{cmdstanr} package enables easy implementation of the Hamiltonian Monte Carlo method \citep{duane1987hybrid, neal2003slice, neal2011mcmc} and its extension, the No-U-Turn sampler \citep{hoffman2014no}. 
While an efficient Bayesian inference algorithm using the Polya-Gamma data augmentation \citep{polson2013bayesian} has been developed for the logit model, no such algorithm has been established for log-log and clog-log models. 
Therefore, in this study, we use \texttt{cmdstanr} for all models. 
Although Bayesian inference via a discrete approximation of the posterior distribution \citep{gelman2013bayesian} could be considered, it is not adopted due to challenges in defining the exploration parameter space and the greater ease of implementation that \texttt{cmdstanr} provides for practitioners. 
A total of 2500 samples are drawn, with the first 500 discarded as burn-in, leaving 2000 samples for posterior estimation of the DLT probability.

\subsection{Results} \label{sec:result}
Table \ref{tab:f_results} shows the MTD selection results for each scenario, comparing the conventional method (PAVA) to the proposed methods using the dose-response models with logit, log-log, and clog-log link functions (logit, log-log, and clog-log). 
The bold values in the table indicate the dose level that should be selected as the MTD for each scenario. 
The proposed methods consistently outperformed the conventional approach, with the logit and clog-log link functions exhibiting particularly strong improvements. 
Notably, in scenario 7, the conventional approach selects the lowest dose as the MTD in 20\% of cases, while our method successfully reduces this occurrence to below 4\%.
As will be discussed in Section \ref{sec:dis}, the proposed method underperformed the conventional approach by a few percentage points in scenario 3. 
However, this scenario is designed without the dose-response model, implying model misspecification. 
Despite the model misspecification, in all other scenarios, the proposed method still outperformed the conventional method by an average of over 5\%. 

As discussed in Section \ref{sec:dr_model}, the log-log link method performs well in scenarios where DLTs frequently occurs at lower doses, while the clog-log link method demonstrates superior performance in scenarios where DLTs frequently occurs at higher doses.
While the log-log model demonstrates the best performance in scenario 4, where the DLT probability at the lowest dose $d_1$ is relatively high at 0.15, it performs poorly in scenario 2, where the MTD was set at the highest dose. 
Typically, in Phase I oncology trials, the lowest dose is determined as one-tenth of the dose causing death in 10 \% of rodents, adjusted for the surface area of the human body. 
Given this, the log-log model may not be the appropriate approach for estimating DLT probabilities in Phase I oncology trials. 

\begin{table}[H]
    \centering
    \caption{The proportion of MTD selections across 1,000 simulations in each scenario for the PAVA-based method and the proposed methods using the logit, log-log, and complementary log-log models. 
    The optimal MTD is highlighted in bold.} \label{tab:f_results} 
    \vspace{2mm}
    \begin{tabular}{cc cccccc} \hline
    \multicolumn{2}{c}{Dose levels} & 1 & 2 & 3 & 4 & 5 & 6 \\ 
    \multicolumn{2}{c}{Doses} & 10 & 20 & 30 & 45 & 60 & 80 \\ \hline 
    scenario 1 & PAVA & 0.9 & 9.2 & 28.1 & $\bm{33.8}$ & 24.6 & 3.4 \\ 
              & logit & 0.0 & 7.3 & 28.3 & $\bm{39.5}$ & 21.0 & 3.9 \\ 
              & log-log & 0.0 & 8.8 & 31.5 & $\bm{41.5}$ & 17.0 & 1.2 \\ 
              & clog-log & 0.0 & 7.1 & 28.0 & $\bm{39.3}$ & 21.6 & 4.0 \\ \hdashline
    scenario 2 & PAVA & 0.0 & 0.7 & 7.4 & 14.6 & 26.6 & $\bm{50.7}$ \\ 
              & logit & 0.0 & 0.0 & 7.4 & 14.7 & 26.3 & $\bm{51.6}$ \\ 
              & log-log & 0.0 & 0.0 & 7.1 & 19.0 & 30.2 & $\bm{43.7}$ \\ 
              & clog-log & 0.0 & 0.0 & 7.0 & 14.9 & 26.0 & $\bm{52.1}$ \\ \hdashline
    scenario 3 & PAVA & 0.0 & 0.2 & 6.9 & 27.4 & $\bm{56.7}$ & 8.8 \\ 
              & logit & 0.0 & 0.0 & 5.1 & 29.8 & $\bm{54.6}$ & 10.5 \\ 
              & log-log & 0.0 & 0.0 & 4.7 & 37.0 & $\bm{52.4}$ & 5.9 \\ 
              & clog-log & 0.0 & 0.0 & 4.8 & 30.4 & $\bm{53.9}$ & 10.9 \\ \hdashline
    scenario 4 & PAVA & 19.6 & $\bm{46.8}$ & 27.2 & 4.5 & 0.8 & 0.0 \\ 
              & logit & 7.6 & $\bm{55.5}$ & 30.5 & 5.2 & 0.1 & 0.0 \\ 
              & log-log & 12.0 & $\bm{58.6}$ & 25.6 & 2.6 & 0.1 & 0.0 \\ 
              & clog-log & 6.3 & $\bm{55.1}$ & 32.0 & 5.2 & 0.3 & 0.0 \\ \hdashline
    scenario 5 & PAVA & 3.1 & 29.2 & $\bm{51.0}$ & 14.1 & 2.4 & 0.1 \\
              & logit & 0.2 & 23.7 & $\bm{57.9}$ & 16.7 & 1.3 & 0.1 \\
              & log-log & 0.4 & 31.3 & $\bm{55.2}$ & 12.2 & 0.8 & 0.0 \\
              & clog-log & 0.1 & 22.3 & $\bm{58.3}$ & 17.6 & 1.5 & 0.1 \\ \hdashline
    scenario 6 & PAVA & 0.2 & 2.6 & 26.3 & $\bm{49.4}$ & 19.6 & 1.9 \\
              & logit & 0.0 & 0.9 & 24.2 & $\bm{55.5}$ & 17.4 & 2.0 \\
              & log-log & 0.0 & 1.0 & 29.5 & $\bm{55.7}$ & 13.2 & 0.6 \\
              & clog-log & 0.0 & 0.9 & 23.7 & $\bm{55.5}$ & 17.8 & 2.1 \\ \hdashline
    scenario 7 & PAVA & 20.0 & $\bm{61.9}$ & 16.9 & 1.1 & 0.0 & 0.0 \\
              & logit & 2.1 & $\bm{73.1}$ & 23.9 & 0.8 & 0.0 & 0.0 \\
              & log-log & 3.8 & $\bm{77.5}$ & 18.3 & 0.3 & 0.0 & 0.0 \\
              & clog-log & 1.2 & $\bm{72.5}$ & 25.4 & 0.8 & 0.0 & 0.0 \\ \hdashline
    scenario 8 & PAVA & 3.1 & 29.5 & $\bm{52.7}$ & 13.3 & 1.3 & 0.0 \\ 
              & logit & 0.2 & 23.8 & $\bm{60.1}$ & 15.1 & 0.7 & 0.0 \\ 
              & log-log & 0.4 & 31.7 & $\bm{57.4}$ & 10.2 & 0.2 & 0.0 \\ 
              & clog-log & 0.1 & 22.3 & $\bm{60.3}$ & 16.5 & 0.7 & 0.0 \\ \hline
    \end{tabular}
\end{table}

Beyond MTD selection accuracy, we also evaluated the safety performance of each method by examining the proportion of overly toxic dose selections (i.e., doses higher than the true MTD). 
Table \ref{tab:overly_toxic_selection} shows these results across all scenarios. 
The log-log link function demonstrated the most conservative behavior, consistently showing the lowest or second-lowest rate of overly toxic dose selections across scenarios 1-8. 
This conservative characteristic was particularly evident in scenarios 4, 5, 6, and 8. 
In contrast, the clog-log link function exhibited more aggressive behavior, often resulting in the highest proportion of overly toxic dose selections among the proposed methods. 
This pattern likely reflects the nature of the clog-log link, which increases toxicity probabilities more gradually at higher doses compared to other link functions. 
The logit link function struck a balanced performance between the conservative log-log and the more aggressive clog-log approaches. Notably, the conventional PAVA method showed substantially higher rates of overly toxic dose selection in several scenarios (1, 4, and 6), underscoring a key limitation of isotonic regression approaches that rely solely on observed data without structural dose-toxicity modeling. 
When considering both MTD selection accuracy and safety performance together, the logit model appears to offer the most well-balanced approach among the evaluated methods. 

\begin{table}[H]
    \centering
    \caption{The proportion of selecting overly toxic doses (doses higher than the true MTD) across 1,000 simulations in each scenario.} \label{tab:overly_toxic_selection}
    \vspace{2mm}
    \begin{tabular}{cc cccc} \hline
    \multicolumn{2}{c}{Method} & PAVA & logit & log-log & clog-log \\ \hline
    scenario 1 &  & 28.0 & 24.9 & 18.2 & 25.6 \\
    scenario 2 &  & 0.0  & 0.0  & 0.0  & 0.0 \\
    scenario 3 &  & 8.8  & 10.5 & 5.9  & 10.9 \\
    scenario 4 &  & 32.5 & 35.8 & 28.3 & 37.5 \\
    scenario 5 &  & 16.6 & 18.1 & 13.0 & 19.2 \\
    scenario 6 &  & 21.5 & 19.4 & 13.8 & 19.9 \\
    scenario 7 &  & 18.0 & 24.7 & 18.6 & 26.2 \\
    scenario 8 &  & 14.6 & 15.8 & 10.4 & 17.2 \\ \hline
    \end{tabular}
\end{table}

\section{Discussion for reference dose and prior distribution} \label{sec:dis} 
The hyperparameters of the prior distribution for the coefficient parameters in dose-response model are influenced by the reference dose. 
Thus, the choice of reference dose may also affect MTD selection.
In the simulations presented in Section \ref{sec:sim}, the reference dose is set to $d^* = d_3$ for all scenarios. 
To investigate the effect of the reference dose, we conduct additional simulations in scenarios 1 to 4, setting $d^* \in \{d_4, d_5, d_6\}$. 
As described in Section \ref{sec:prior}, the selection of the reference dose influences the hyperparameters of the prior distribution for the coefficient parameters. 
Specifically, when the reference dose is set to $d_4$, $d_5$, or $d_6$, the prior distributions corresponding to each link function are 
\begin{equation*}
    \begin{split}
        \mbox{logit}   &: \beta_0 \sim N(-0.542,1.197), \quad \beta_1 \sim N(0.407,0.500), \\
        \mbox{log-log} &: \beta_0 \sim N(0.272,0.773), \quad \beta_1 \sim N(-0.242,0.500), \\
        \mbox{clog-log}&: \beta_0 \sim N(-0.896,0.928), \quad \beta_1 \sim N(0.309,0.571),
    \end{split}
\end{equation*}
\begin{equation*}
    \begin{split}
        \mbox{logit}   &: \beta_0 \sim N(0.084,1.540), \quad \beta_1 \sim N(0.822,0.500), \\
        \mbox{log-log} &: \beta_0 \sim N(0.444,0.894), \quad \beta_1 \sim N(-0.004,0.500), \\
        \mbox{clog-log}&: \beta_0 \sim N(-0.431,0.776), \quad \beta_1 \sim N(0.221,0.500),
    \end{split}
\end{equation*}
\begin{equation*}
    \begin{split}
        \mbox{logit}   &: \beta_0 \sim N(0.371,1.209), \quad \beta_1 \sim N(0.600,0.500), \\
        \mbox{log-log} &: \beta_0 \sim N(0.673,0.712), \quad \beta_1 \sim N(-0.054,0.594), \\
        \mbox{clog-log}&: \beta_0 \sim N(-0.294,0.857), \quad \beta_1 \sim N(0.235,0.500),
    \end{split}
\end{equation*}
respectively. 

Tables \ref{tab:refd4}, \ref{tab:refd5}, and \ref{tab:refd6} show the results obtained using logit, log-log, and clog-log link functions, respectively. 
In scenario 2, the accuracy of MTD selection declines as the reference dose increased, while in Scenario 3, it shows improvement. 
For the remaining scenarios, the reference dose has little impact. 
A common feature of scenarios 2 and 3, which exhibit significant variability in results, is that the MTDs are set at higher dose levels. 
Table \ref{tab:numberOfNandY} shows the averages of $n_j$ and $m_j$ over 1000 datasets for each dose level in every scenario. 
Note that $n_j$ and $m_j$ remain unchanged regardless of the reference dose. 
In scenarios 2 and 3, the number of DLTs, $m_j$, is lower than in other scenarios. 
This is because, in trials where the MTD is set at a higher dose, there is typically a lower number of DLTs at higher doses. 
Generally, in scenarios where an insufficient number of DLTs are observed, such as when the MTD is near the highest dose, the influence of the likelihood on DLT probability estimation decreases, making the prior distribution relatively more influential. 
Consequently, the dependence of the prior distribution on the reference dose setting may have led to increased variability in the results of scenarios 2 and 3, particularly due to a few number of DLTs at higher doses. 
In contrast, in scenario 4, where the number of DLT occurrences is larger, the influence of the reference dose selection is minimal. 
Nevertheless, since cases where the MTD is set at the highest dose are extremely rare in Phase I oncology trials, the observed instability in scenarios 2 and 3 is not of major concern. 

Given that the influence of reference dose selection is most pronounced in scenarios where the MTD is at higher dose levels (which are extremely rare in Phase I oncology trials), we recommend a conservative approach of selecting the reference dose at or near the middle of the dose range (e.g., $d_3$ or $d_4$ in a 6-dose study). 
This minimizes potential instability while maintaining good performance across most realistic scenarios. 
For investigators concerned about reference dose sensitivity, conducting a sensitivity analysis with different reference dose choices during the trial planning phase would provide additional assurance about the robustness of the design. 

\begin{table}[H]
    \caption{The proportion of MTD selections across 1,000 simulations in each scenario for the proposed method using the logit model when the reference dose $d^*$ is $d_4$, $d_5$, or $d_6$. 
    The optimal MTD is highlighted in bold.} 
    \label{tab:refd4}
    \centering
    \vspace{2mm}
    \begin{tabular}{cc cccccc} \hline
    \multicolumn{2}{c}{Dose levels} & 1 & 2 & 3 & 4 & 5 & 6 \\ 
    \multicolumn{2}{c}{Doses} & 10 & 20 & 30 & 45 & 60 & 80 \\ \hline 
    scenario1 & $d^*=d_4$ & 0.0 & 6.5 & 27.3 & $\bm{43.4}$ & 20.7 & 2.1 \\ 
              & $d^*=d_5$ & 0.0 & 7.3 & 27.8 & $\bm{42.5}$ & 21.0 & 1.4 \\ 
              & $d^*=d_6$ & 0.0 & 6.4 & 26.6 & $\bm{43.9}$ & 21.8 & 1.3 \\ \hdashline
    scenario2 & $d^*=d_4$ & 0.0 & 0.0 & 5.9 & 17.0 & 30.2 & $\bm{46.9}$ \\ 
              & $d^*=d_5$ & 0.0 & 0.0 & 6.0 & 18.1 & 31.9 & $\bm{44.0}$ \\ 
              & $d^*=d_6$ & 0.0 & 0.0 & 5.2 & 17.8 & 33.4 & $\bm{43.6}$ \\ \hdashline
    scenario3 & $d^*=d_4$ & 0.0 & 0.0 & 3.0 & 33.7 & $\bm{55.7}$ & 7.6 \\ 
              & $d^*=d_5$ & 0.0 & 0.0 & 3.6 & 33.2 & $\bm{57.8}$ & 5.4 \\ 
              & $d^*=d_6$ & 0.0 & 0.0 & 2.6 & 31.8 & $\bm{60.4}$ & 5.2 \\ \hdashline
    scenario4 & $d^*=d_4$ & 8.4 & $\bm{56.2}$ & 29.1 & 5.1 & 0.1 & 0.0 \\ 
              & $d^*=d_5$ & 10.8 & $\bm{55.2}$ & 27.9 & 4.9 & 0.1 & 0.0 \\ 
              & $d^*=d_6$ & 8.8 & $\bm{55.8}$ & 29.0 & 5.2 & 0.1 & 0.0 \\ \hline
    \end{tabular}
\end{table}

\begin{table}[H]
    \caption{The proportion of MTD selections across 1,000 simulations in each scenario for the proposed method using the log-log model when the reference dose $d^*$ is $d_4$, $d_5$, or $d_6$. 
    The optimal MTD is highlighted in bold.} 
    \label{tab:refd5} 
    \centering 
    \vspace{2mm}
    \begin{tabular}{cc cccccc} \hline
    \multicolumn{2}{c}{Dose levels} & 1 & 2 & 3 & 4 & 5 & 6 \\ 
    \multicolumn{2}{c}{Doses} & 10 & 20 & 30 & 45 & 60 & 80 \\ \hline 
    scenario 1 & $d^*=d_4$ & 0.0 & 7.0 & 30.0 & $\bm{44.1}$ & 17.6 & 1.3 \\ 
              & $d^*=d_5$ & 0.0 & 7.0 & 30.8 & $\bm{44.3}$ & 16.9 & 1.0 \\ 
              & $d^*=d_6$ & 0.0 & 6.6 & 30.7 & $\bm{45.3}$ & 16.7 & 0.7 \\ \hdashline
    scenario 2 & $d^*=d_4$ & 0.0 & 0.0 & 5.9 & 19.4 & 30.7 & $\bm{47.4}$ \\ 
              & $d^*=d_5$ & 0.0 & 0.0 & 6.0 & 20.4 & 31.8 & $\bm{41.8}$ \\ 
              & $d^*=d_6$ & 0.0 & 0.0 & 5.8 & 21.0 & 34.3 & $\bm{38.9}$ \\ \hdashline
    scenario 3 & $d^*=d_4$ & 0.0 & 0.0 & 3.7 & 36.8 & $\bm{53.7}$ & 5.8 \\ 
              & $d^*=d_5$ & 0.0 & 0.0 & 4.0 & 36.3 & $\bm{54.6}$ & 5.1 \\ 
              & $d^*=d_6$ & 0.0 & 0.0 & 3.6 & 36.1 & $\bm{56.3}$ & 4.0 \\ \hdashline
    scenario 4 & $d^*=d_4$ & 12.3 & $\bm{58.2}$ & 25.3 & 3.0 & 0.1 & 0.0 \\ 
              & $d^*=d_5$ & 12.2 & $\bm{57.8}$ & 25.9 & 2.9 & 0.1 & 0.0 \\ 
              & $d^*=d_6$ & 11.0 & $\bm{58.6}$ & 26.2 & 3.0 & 0.1 & 0.0 \\ \hline
    \end{tabular}
\end{table}

\begin{table}[H]
    \caption{The proportion of MTD selections across 1,000 simulations in each scenario for the proposed method using the clog-log model when the reference dose $d^*$ is $d_4$, $d_5$, or $d_6$. 
    The optimal MTD is highlighted in bold.} 
    \label{tab:refd6} 
    \centering
    \vspace{2mm}
    \begin{tabular}{cc cccccc} \hline 
    \multicolumn{2}{c}{Dose levels} & 1 & 2 & 3 & 4 & 5 & 6 \\ 
    \multicolumn{2}{c}{Doses} & 10 & 20 & 30 & 45 & 60 & 80 \\ \hline 
    scenario 1 & $d^*=d_4$ & 0.0 & 6.3 & 27.2 & $\bm{42.7}$ & 21.5 & 2.3 \\ 
              & $d^*=d_5$ & 0.0 & 5.9 & 26.3 & $\bm{45.1}$ & 21.7 & 1.0 \\ 
              & $d^*=d_6$ & 0.0 & 5.8 & 24.0 & $\bm{44.6}$ & 23.5 & 2.1 \\ \hdashline
    scenario 2 & $d^*=d_4$ & 0.0 & 0.0 & 6.3 & 16.7 & 29.6 & $\bm{52.1}$ \\ 
              & $d^*=d_5$ & 0.0 & 0.0 & 5.0 & 17.9 & 35.8 & $\bm{41.3}$ \\ 
              & $d^*=d_6$ & 0.0 & 0.0 & 4.4 & 17.2 & 30.3 & $\bm{48.1}$ \\ \hdashline
    scenario 3 & $d^*=d_4$ & 0.0 & 0.0 & 3.6 & 32.8 & $\bm{56.5}$ & 7.1 \\ 
              & $d^*=d_5$ & 0.0 & 0.0 & 2.5 & 34.4 & $\bm{58.6}$ & 4.5 \\ 
              & $d^*=d_6$ & 0.0 & 0.0 & 2.0 & 28.0 & $\bm{61.5}$ & 8.5 \\ \hdashline
    scenario 4 & $d^*=d_4$ & 7.2 & $\bm{55.1}$ & 31.2 & 5.3 & 0.1 & 0.0 \\ 
              & $d^*=d_5$ & 7.1 & $\bm{55.8}$ & 30.5 & 5.4 & 0.1 & 0.0 \\ 
              & $d^*=d_6$ & 7.2 & $\bm{54.9}$ & 30.7 & 6.0 & 0.1 & 0.0 \\ \hline
    \end{tabular}
\end{table}

\begin{table}[H]
    \caption{The averages of allocated patients, $n_j$, and DLTs, $m_j$, over 1,000 datasets for each dose level in every scenario. 
    The optimal MTD is highlighted in bold. } 
    \label{tab:numberOfNandY} 
    \centering 
    \vspace{2mm} 
    \begin{tabular}{cc cccccc} \hline 
    \multicolumn{2}{c}{Dose levels} & 1 & 2 & 3 & 4 & 5 & 6 \\ 
    \multicolumn{2}{c}{Doses} & 10 & 20 & 30 & 45 & 60 & 80 \\ \hline 
    scenario1 & $n_j$ & 3.915 & 7.500 & 9.990 & $\bm{8.460}$ & 4.707 & 1.428 \\ 
              & $m_j$ & 0.066 & 1.136 & 1.950 & $\bm{2.561}$ & 1.630 & 0.775 \\ \hdashline
    scenario2 & $n_j$ & 3.084 & 3.987 & 6.597 & 7.248 & 6.840 & $\bm{8.244}$ \\ 
              & $m_j$ & 0.025 & 0.168 & 0.899 & 1.370 & 1.459 & $\bm{2.443}$ \\ \hdashline
    scenario3 & $n_j$ & 3.078 & 3.534 & 6.096 & 9.513 & $\bm{10.140}$ & 3.639 \\ 
              & $m_j$ & 0.025 & 0.110 & 0.596 & 1.894 & $\bm{3.070}$ & 1.960 \\ \hdashline
    scenario4 & $n_j$ & 11.196 & $\bm{14.037}$ & 7.938 & 2.133 & 0.339 & 0.027 \\ 
              & $m_j$ & 1.642 & $\bm{4.218}$ & 2.874 & 1.062 & 0.177 & 0.021 \\ \hdashline
    scenario5 & $n_j$ & 5.559 & 11.868 & $\bm{12.327}$ & 5.064 & 1.059 & 0.093 \\
              & $m_j$ & 0.419 & 2.256 & $\bm{3.694}$ & 2.253 & 0.546 & 0.060 \\ \hdashline
    scenario6 & $n_j$ & 3.471 & 5.280 & 10.116 & $\bm{10.947}$ & 5.079 & 1.107 \\
              & $m_j$ & 0.096 & 0.494 & 1.656 & $\bm{3.306}$ & 2.127 & 0.611 \\ \hdashline
    scenario7 & $n_j$ & 10.581 & $\bm{17.337}$ & 6.873 & 1.089 & 0.090 & 0.000 \\
              & $m_j$ & 0.906 & $\bm{5.224}$ & 3.072 & 0.629 & 0.062 & 0.000 \\ \hdashline
    scenario8 & $n_j$ & 5.559 & 11.895 & $\bm{12.624}$ & 5.022 & 0.816 & 0.054 \\ 
              & $m_j$ & 0.419 & 2.258 & $\bm{3.792}$ & 2.307 & 0.479 & 0.039 \\ \hline
    \end{tabular}
\end{table}

The adoption of a dose-response model that excludes a reference dose, such as 
\begin{equation} \label{eq:nmodel}
    g(\pi(d_j)) = \beta_0 + \exp(\beta_1)\log(d_j), 
\end{equation}
could potentially address the variability in MTD selection results caused by reference dose settings.
However, this comes at the cost of sacrificing the interpretability of coefficient parameters, as discussed in Section \ref{sec:dr_model}. 
Moreover, since the reference dose in our dose-response model \eqref{eq:drmodel} adjusts the scale of the dose $\{d_j\}$, eliminating it may cause numerical instability due to increased variability in the coefficient  $\log(d_j)$ for $\beta_1$. 
The specification of prior distributions in the dose-response model \eqref{eq:nmodel} without the reference dose and the evaluation of MTD selection in comparison with the proposed approach incorporating the dose-response model \eqref{eq:drmodel} with the reference dose will be left as future research challenges.

\section{Conclusion and remarks} \label{sec:concl} 
In this study, we propose replacing PAVA-based isotonic regression with a Bayesian dose-response model to improve the accuracy of MTD selection in model-assisted designs for Phase I oncology trials. 
The effectiveness of this approach is validated through extensive simulations. 
As described in Section \ref{sec:pava}, PAVA adjusts independently estimated DLT probabilities at each dose level that violate the assumption of monotonicity using posterior variances. 
Therefore, in Phase I oncology trials with small sample sizes, the estimated DLT probabilities tend to be unstable. 
On the other hand, our approach with the dose-response model is expected to achieve higher MTD selection accuracy than the PAVA-based approach, as it allows for the estimation of DLT probabilities at dose levels with small sample sizes by borrowing information from other dose levels under the assumption of monotonicity. 
Indeed, our simulation study confirms that the proposed method outperforms the conventional method in most scenarios. 
While the conventional method performs slightly better in some cases, these occur when the MTD is set at a high dose level, which are rare in Phase I oncology trials. 
Hence, we do not consider this to be a critical issue from our perspective. 
Moreover, we emphasize that the proposed method still outperforms the conventional method even in scenarios where the assumed dose-response model is misspecified. 

In our dose-response model, we explore the use of three link functions: logit, log-log, and clog-log. 
As detailed in Section \ref{sec:dr_model}, each of these link functions introduces distinct properties to the model. 
Our simulations confirm that the clog-log link enhances MTD selection accuracy when DLT probabilities are low at lower doses, while the log-log link is more effective when DLT probabilities are high at higher doses. 
Meanwhile, the logit link demonstrates relatively stable performance across all scenarios while still outperforming the conventional method. 
Ideally, the selection of a link function should be based on discussions with practitioners to ensure it accurately reflects practical dose-response relationships. 
When no prior information is available for selecting a link function, one-parameter or two-parameter link functions \citep{aranda1981two, stukel1988generalized, chen1999new} may serve as viable data-driven alternatives. 
However, incorporating additional parameters in small sample settings may compromise the stability of the estimates. 
The exploration of these alternative link functions is reserved for future research. 

Finally, we aim to investigate whether using dose-response models for DLT probability estimation can improve MTD selection accuracy in model-assisted designs beyond the BOIN design and assess whether dose-response models can be adopted as a standard method in place of isotonic regression. 

\def\thesection{Appendix}
\def\thesubsection{A.\arabic{subsection}}
\section{}
\setcounter{equation}{0}
\setcounter{table}{0}
\setcounter{figure}{0}
\def\theequation{A.\arabic{equation}}
\def\thethm{A.\arabic{thm}}
\def\thelem{A.\arabic{lem}}
\renewcommand{\thetable}{A.\arabic{table}}
\renewcommand{\thefigure}{A.\arabic{figure}}

\subsection{Sensitivity Analyses for Prior Specification}
To assess the robustness of our proposed method to prior specification choices, we conducted sensitivity analyses using the same simulation settings as described in Section 4.1 of the main manuscript. 
All simulations were conducted under identical conditions to the main study: 3 patients per cohort, 12 cohorts (36 participants total), six dose levels (10, 20, 30, 45, 60, 80) mg, target DLT probability $\phi = 0.3$, and the same eight DLT probability scenarios (for details, see Table 3 in the main manuscript). 
We applied the Bayesian dose-response model,
\begin{equation*} 
    g(\pi(d_j)) = \beta_0 + \exp(\beta_1)\log\left(\frac{d_j}{d^*}\right), 
\end{equation*}
where $g(\cdot)$ is the link function, $\beta_0$ and $\beta_1$ are coefficient parameters, and $\pi(d_j)$ is the DLT probability for dose $d_j$. 
The $d^*$, referred to as the reference dose, is selected from $\{d_j\}$ (e.g., $d^*=d_3$). 
We examined two types of prior specification using logit, log-log, and complementary log-log link functions: 
(1) alternative specifications of $p_1$ and $p_J$ values that differ from our BOIN-aligned defaults, and 
(2) large variance normal priors for coefficient parameters that lead to inappropriate U-shaped prior distributions for intermediate doses. 
Note that Bayesian inference was performed using the same computational approach as described in Section 4.2, employing the \texttt{cmdstanr} package with 2500 samples (500 burn-in, 2000 for posterior estimation). 

\subsubsection{Alternative specifications of $p_1$ and $p_J$ values} 
We examined the impact of alternative specifications for the probability values $p_1 = P\{\pi(d_1) > 0.3\}$ and $p_J = P\{\pi(d_J) \leq 0.3\}$ that differ from our BOIN-aligned defaults used in the main analysis. 
As described in Section 3.2 of the main manuscript, our default specification sets $p_1 = 0.05$ for the lowest dose and $p_J = 0.21$ (corresponding to $P\{\pi(d_J) \leq 0.1\} = 0.05$) for the highest dose, which aligns with BOIN's safety mechanisms and addresses the dual concerns of controlling unacceptable toxicity risk while avoiding underdosing. 
For $p_1$, we considered four alternative values: $p_1 = 0.09$ (corresponding to $P\{\pi(d_1) > \phi + \Delta_U\} = 0.05$, where $\phi + \Delta_U = 0.3585$ represents the BOIN escalation boundary); $p_1 = 0.2$ (representing a moderately conservative approach); $p_1 = 0.31$ \citep[corresponding to $P\{\pi(d_1) > 0.6\} = 0.05$, which is the default setting of][]{neuenschwander2008critical}; and $p_1 = 0.7$ (corresponding to $\pi(d_1) \sim \text{Beta}(1,1)$, representing a uniform prior). 
For $p_J$, we examined three alternative values: $p_J = 0.08$ (corresponding to $P\{\pi(d_J) \leq \phi - \Delta_L\} = 0.05$, where $\phi - \Delta_L = 0.2365$ represents the BOIN de-escalation boundary), $p_J = 0.11$ \citep[corresponding to $P\{\pi(d_J) \leq 0.2\} = 0.05$, the default setting from][]{neuenschwander2008critical}, and $p_J = 0.3$ (corresponding to $\pi(d_J) \sim \text{Beta}(1,1)$). 
To isolate the individual effects of each parameter, we conducted separate sensitivity analyses, that is, when examining $p_1$ sensitivity, we fixed $p_J = 0.21$ (our default); when examining $p_J$ sensitivity, we fixed $p_1 = 0.05$ (our default). 
For each alternative specification, we followed the same procedure described in Section 3.2 to derive the corresponding prior distributions for the coefficient parameters $\beta_0$ and $\beta_1$ through the optimization process that minimizes the discrepancy between target quantiles and model-implied quantiles. 

Tables \ref{tab:p1-1}-\ref{tab:pJ-3} present the MTD selection results under alternative specifications of $p_1$ and $p_J$ values across all eight scenarios. 
Comparing these results with Table 4 in the main manuscript (our default specification: $p_1 = 0.05$, $p_J = 0.21$) reveals several important sensitivity patterns. 

For $p_1$ sensitivity analysis (Tables \ref{tab:p1-1}-\ref{tab:p1-4}, with $p_J$ fixed at 0.21), the results demonstrate that our method is generally robust to moderate variations in $p_1$ specification. 
The alternative specifications $p_1 = 0.09$, $p_1 = 0.2$, and $p_1 = 0.31$ show minimal differences compared to our default specification ($p_1 = 0.05$), with the MTD selection accuracy remaining largely stable in most scenarios. 
However, the extreme specification $p_1 = 0.7$ (corresponding to a uniform prior) shows notable performance deterioration, particularly in scenarios with intermediate MTDs. 
In Scenarios 1, 3, 4, 5, 6, 7, and 8, the logit model shows an average decrease of approximately 8 percentage points in the correct MTD selection rate. 
More concerning, some link functions experience reductions exceeding about 20 percentage points. 
For example, in Scenario 8, the log-log model decreases from 57.4\% to 36.8\% (a 20.6 percentage point reduction), and the clog-log model decreases from 60.3\% to 40.8\% (a 19.5 percentage point reduction). 
In contrast, Scenario 2 (where the MTD is at the highest dose level) shows improved performance with $p_1 = 0.7$. 
The logit model increases from 51.6\% to 61.1\% (a 9.5 percentage point improvement), with even larger improvements observed for the log-log model (43.7\% to 64.0\%, a 20.3 percentage point increase). 
This improvement likely occurs because the specification $p_1 = 0.7$ implies that the lowest dose is expected to have relatively high toxicity. 
Through the dose-response model, this translates to a prior distribution that accommodates higher DLT probabilities at higher doses. 
This may lead to more aggressive dose selection that favors higher doses, which aligns well with Scenario 2 where the true MTD is at the highest dose level. 

For $p_J$ sensitivity analysis (Tables \ref{tab:pJ-1}-\ref{tab:pJ-3}, with $p_1$ fixed at 0.05), the results demonstrate that our method shows greater robustness to variations in $p_J$ specification compared to $p_1$ variations. 
All alternative specifications show minimal impact on MTD selection accuracy across most scenarios, with differences typically within approximately 1-3 percentage points compared to our default specification ($p_J = 0.21$). 
However, one notable exception occurs with $p_J = 0.08$ in Scenario 2 (MTD at the highest dose level), where the logit model shows a substantial decrease from 51.6\% to 43.6\% (an 8.0 percentage point reduction). 
This deterioration likely occurs because the more conservative specification $p_J = 0.08$ implies lower expected toxicity at the highest dose, which may discourage selection of higher doses even when they represent the true MTD. 

The relatively lower sensitivity to $p_J$ variations compared to $p_1$ variations suggests that the specification of prior beliefs about toxicity at the lowest dose level has a more substantial impact on overall model behavior than beliefs about the highest dose level. 
This finding aligns with the clinical importance of establishing appropriate safety constraints at the lowest dose level in phase I trials. 
Across both $p_1$ and $p_J$ sensitivity analyses, the patterns remain consistent across all three link functions (logit, log-log, and complementary log-log), demonstrating that the sensitivity patterns are inherent to the prior specification approach rather than dependent on the choice of link function.

\begin{table}[H]
    \centering
    \caption{MTD selection results for $p_1 = 0.09$ with $p_J = 0.21$. Prior distributions: logit: $\beta_0 \sim N(-1.067, 1.179)$, $\beta_1 \sim N(0.219, 0.898)$; log-log: $\beta_0 \sim N(-0.196, 1.217)$, $\beta_1 \sim N(0.328, 0.569)$; clog-log: $\beta_0 \sim N(-1.255, 0.932)$, $\beta_1 \sim N(0.092, 0.642)$.} \label{tab:p1-1} 
    \begin{tabular}{cc cccccc} \hline
    \multicolumn{2}{c}{Dose levels} & 1 & 2 & 3 & 4 & 5 & 6 \\ 
    \multicolumn{2}{c}{Doses} & 10 & 20 & 30 & 45 & 60 & 80 \\ \hline 
    scenario 1 & PAVA & 0.9 & 9.2 & 28.1 & $\bm{33.8}$ & 24.6 & 3.4 \\ 
              & logit & 0.0 & 7.4 & 29.4 & $\bm{39.0}$ & 19.9 & 4.3 \\ 
              & log-log & 0.0 & 9.5 & 32.9 & $\bm{40.4}$ & 16.2 & 1.0 \\ 
              & clog-log & 0.0 & 7.4 & 28.5 & $\bm{39.8}$ & 20.7 & 3.6 \\ \hdashline
    scenario 2 & PAVA & 0.0 & 0.7 & 7.4 & 14.6 & 26.6 & $\bm{50.7}$ \\ 
              & logit & 0.0 & 0.0 & 7.5 & 14.3 & 26.0 & $\bm{52.2}$ \\ 
              & log-log & 0.0 & 0.0 & 7.6 & 19.6 & 31.0 & $\bm{41.8}$ \\ 
              & clog-log & 0.0 & 0.0 & 6.9 & 15.1 & 26.9 & $\bm{51.1}$ \\ \hdashline
    scenario 3 & PAVA & 0.0 & 0.2 & 6.9 & 27.4 & $\bm{56.7}$ & 8.8 \\ 
              & logit & 0.0 & 0.0 & 5.4 & 31.4 & $\bm{51.5}$ & 11.7 \\ 
              & log-log & 0.0 & 0.0 & 5.7 & 36.5 & $\bm{52.9}$ & 4.9 \\ 
              & clog-log & 0.0 & 0.0 & 4.5 & 32.2 & $\bm{52.8}$ & 10.5 \\ \hdashline
    scenario 4 & PAVA & 19.6 & $\bm{46.8}$ & 27.2 & 4.5 & 0.8 & 0.0 \\ 
              & logit & 8.2 & $\bm{56.4}$ & 29.2 & 5.0 & 0.1 & 0.0 \\ 
              & log-log & 15.7 & $\bm{56.8}$ & 23.8 & 2.5 & 0.1 & 0.0 \\ 
              & clog-log & 7.3 & $\bm{56.4}$ & 30.0 & 5.1 & 0.1 & 0.0 \\ \hdashline
    scenario 5 & PAVA & 3.1 & 29.2 & $\bm{51.0}$ & 14.1 & 2.4 & 0.1 \\
              & logit & 0.2 & 24.3 & $\bm{59.4}$ & 14.7 & 1.2 & 0.1 \\
              & log-log & 1.2 & 32.9 & $\bm{53.3}$ & 11.7 & 0.8 & 0.0 \\
              & clog-log & 0.2 & 23.9 & $\bm{58.8}$ & 15.7 & 1.2 & 0.1 \\ \hdashline
    scenario 6 & PAVA & 0.2 & 2.6 & 26.3 & $\bm{49.4}$ & 19.6 & 1.9 \\
              & logit & 0.0 & 0.9 & 26.1 & $\bm{53.8}$ & 17.0 & 2.2 \\
              & log-log & 0.0 & 1.4 & 30.9 & $\bm{54.6}$ & 12.5 & 0.6 \\
              & clog-log & 0.0 & 0.9 & 24.5 & $\bm{55.5}$ & 17.4 & 1.7 \\ \hdashline
    scenario 7 & PAVA & 20.0 & $\bm{61.9}$ & 16.9 & 1.1 & 0.0 & 0.0 \\
              & logit & 2.4 & $\bm{74.0}$ & 22.9 & 0.6 & 0.0 & 0.0 \\
              & log-log & 6.7 & $\bm{76.4}$ & 16.5 & 0.3 & 0.0 & 0.0 \\
              & clog-log & 1.8 & $\bm{73.6}$ & 23.8 & 0.7 & 0.0 & 0.0 \\ \hdashline
    scenario 8 & PAVA & 3.1 & 29.5 & $\bm{52.7}$ & 13.3 & 1.3 & 0.0 \\ 
              & logit & 0.2 & 24.5 & $\bm{61.5}$ & 13.2 & 0.5 & 0.0 \\ 
              & log-log & 1.2 & 33.4 & $\bm{55.3}$ & 9.8 & 0.2 & 0.0 \\ 
              & clog-log & 0.2 & 24.0 & $\bm{61.0}$ & 14.2 & 0.5 & 0.0 \\ \hline
    \end{tabular}
\end{table}

\begin{table}[H]
    \centering
    \caption{MTD selection results for $p_1 = 0.20$ with $p_J = 0.21$. Prior distributions: logit: $\beta_0 \sim N(-0.945, 1.402)$, $\beta_1 \sim N(0.103, 0.742)$; log-log: $\beta_0 \sim N(0.008, 0.867)$, $\beta_1 \sim N(-0.160, 0.500)$; clog-log: $\beta_0 \sim N(-1.121, 0.937)$, $\beta_1 \sim N(-0.165, 0.661)$.} \label{tab:p1-2} 
    \begin{tabular}{cc cccccc} \hline
    \multicolumn{2}{c}{Dose levels} & 1 & 2 & 3 & 4 & 5 & 6 \\ 
    \multicolumn{2}{c}{Doses} & 10 & 20 & 30 & 45 & 60 & 80 \\ \hline 
    scenario 1 & PAVA & 0.9 & 9.2 & 28.1 & $\bm{33.8}$ & 24.6 & 3.4 \\ 
              & logit & 0.0 & 7.6 & 28.2 & $\bm{39.4}$ & 20.3 & 4.5 \\ 
              & log-log & 0.0 & 8.5 & 30.4 & $\bm{42.2}$ & 17.0 & 1.9 \\ 
              & clog-log & 0.0 & 7.4 & 26.6 & $\bm{40.4}$ & 21.0 & 4.6 \\ \hdashline
    scenario 2 & PAVA & 0.0 & 0.7 & 7.4 & 14.6 & 26.6 & $\bm{50.7}$ \\ 
              & logit & 0.0 & 0.0 & 6.9 & 14.3 & 26.0 & $\bm{52.8}$ \\ 
              & log-log & 0.0 & 0.0 & 6.5 & 18.6 & 29.6 & $\bm{45.3}$ \\ 
              & clog-log & 0.0 & 0.0 & 6.4 & 15.0 & 25.5 & $\bm{53.1}$ \\ \hdashline
    scenario 3 & PAVA & 0.0 & 0.2 & 6.9 & 27.4 & $\bm{56.7}$ & 8.8 \\ 
              & logit & 0.0 & 0.0 & 4.4 & 30.0 & $\bm{53.5}$ & 12.1 \\ 
              & log-log & 0.0 & 0.0 & 4.1 & 36.7 & $\bm{51.6}$ & 7.6 \\ 
              & clog-log & 0.0 & 0.0 & 4.2 & 30.4 & $\bm{53.3}$ & 12.1 \\ \hdashline
    scenario 4 & PAVA & 19.6 & $\bm{46.8}$ & 27.2 & 4.5 & 0.8 & 0.0 \\ 
              & logit & 9.9 & $\bm{56.0}$ & 28.1 & 4.8 & 0.1 & 0.0 \\ 
              & log-log & 12.8 & $\bm{58.7}$ & 24.5 & 2.8 & 0.1 & 0.0 \\ 
              & clog-log & 8.3 & $\bm{56.2}$ & 29.1 & 5.2 & 0.1 & 0.0 \\ \hdashline
    scenario 5 & PAVA & 3.1 & 29.2 & $\bm{51.0}$ & 14.1 & 2.4 & 0.1 \\
              & logit & 0.2 & 25.4 & $\bm{58.0}$ & 14.9 & 1.3 & 0.1 \\
              & log-log & 0.5 & 31.5 & $\bm{55.1}$ & 12.0 & 0.8 & 0.0 \\
              & clog-log & 0.2 & 24.2 & $\bm{57.7}$ & 16.4 & 1.3 & 0.1 \\ \hdashline
    scenario 6 & PAVA & 0.2 & 2.6 & 26.3 & $\bm{49.4}$ & 19.6 & 1.9 \\
              & logit & 0.0 & 0.9 & 24.6 & $\bm{54.1}$ & 18.2 & 2.2 \\
              & log-log & 0.0 & 0.9 & 27.8 & $\bm{56.8}$ & 13.7 & 0.8 \\
              & clog-log & 0.0 & 0.9 & 21.7 & $\bm{57.0}$ & 18.2 & 2.2 \\ \hdashline
    scenario 7 & PAVA & 20.0 & $\bm{61.9}$ & 16.9 & 1.1 & 0.0 & 0.0 \\
              & logit & 2.9 & $\bm{75.2}$ & 21.1 & 0.7 & 0.0 & 0.0 \\
              & log-log & 4.3 & $\bm{77.8}$ & 17.5 & 0.3 & 0.0 & 0.0 \\
              & clog-log & 2.4 & $\bm{74.5}$ & 22.1 & 0.9 & 0.0 & 0.0 \\ \hdashline
    scenario 8 & PAVA & 3.1 & 29.5 & $\bm{52.7}$ & 13.3 & 1.3 & 0.0 \\ 
              & logit & 0.2 & 25.8 & $\bm{59.9}$ & 13.4 & 0.6 & 0.0 \\ 
              & log-log & 0.5 & 32.0 & $\bm{57.0}$ & 10.2 & 0.2 & 0.0 \\ 
              & clog-log & 0.2 & 24.4 & $\bm{59.9}$ & 14.7 & 0.7 & 0.0 \\ \hline
    \end{tabular}
\end{table}

\begin{table}[H]
    \centering
    \caption{MTD selection results for $p_1 = 0.31$ with $p_J = 0.21$. Prior distributions: logit: $\beta_0 \sim N(-0.599, 1.515)$, $\beta_1 \sim N(-0.124, 0.500)$; log-log: $\beta_0 \sim N(0.143, 0.824)$, $\beta_1 \sim N(-1.176, 1.470)$; clog-log: $\beta_0 \sim N(-0.991, 1.020)$, $\beta_1 \sim N(-0.447, 0.542)$.} \label{tab:p1-3} 
    \begin{tabular}{cc cccccc} \hline
    \multicolumn{2}{c}{Dose levels} & 1 & 2 & 3 & 4 & 5 & 6 \\ 
    \multicolumn{2}{c}{Doses} & 10 & 20 & 30 & 45 & 60 & 80 \\ \hline 
    scenario 1 & PAVA & 0.9 & 9.2 & 28.1 & $\bm{33.8}$ & 24.6 & 3.4 \\ 
              & logit & 0.0 & 6.8 & 26.1 & $\bm{39.3}$ & 22.3 & 5.5 \\ 
              & log-log & 0.0 & 8.8 & 29.0 & $\bm{37.4}$ & 19.6 & 5.2 \\ 
              & clog-log & 0.0 & 6.8 & 24.3 & $\bm{40.2}$ & 22.8 & 5.9 \\ \hdashline
    scenario 2 & PAVA & 0.0 & 0.7 & 7.4 & 14.6 & 26.6 & $\bm{50.7}$ \\ 
              & logit & 0.0 & 0.0 & 5.3 & 15.5 & 23.4 & $\bm{55.8}$ \\ 
              & log-log & 0.0 & 0.2 & 7.6 & 13.8 & 23.8 & $\bm{54.6}$ \\ 
              & clog-log & 0.0 & 0.0 & 5.1 & 15.6 & 22.9 & $\bm{56.4}$ \\ \hdashline
    scenario 3 & PAVA & 0.0 & 0.2 & 6.9 & 27.4 & $\bm{56.7}$ & 8.8 \\ 
              & logit & 0.0 & 0.0 & 3.0 & 28.7 & $\bm{51.4}$ & 16.9 \\ 
              & log-log & 0.0 & 0.0 & 6.5 & 29.8 & $\bm{49.6}$ & 14.1 \\ 
              & clog-log & 0.0 & 0.0 & 2.8 & 28.7 & $\bm{50.7}$ & 17.8 \\ \hdashline
    scenario 4 & PAVA & 19.6 & $\bm{46.8}$ & 27.2 & 4.5 & 0.8 & 0.0 \\ 
              & logit & 11.3 & $\bm{56.7}$ & 25.9 & 4.9 & 0.1 & 0.0 \\ 
              & log-log & 16.0 & $\bm{55.0}$ & 23.8 & 3.7 & 0.4 & 0.0 \\ 
              & clog-log & 10.6 & $\bm{55.9}$ & 27.2 & 5.0 & 0.2 & 0.0 \\ \hdashline
    scenario 5 & PAVA & 3.1 & 29.2 & $\bm{51.0}$ & 14.1 & 2.4 & 0.1 \\
              & logit & 0.3 & 26.0 & $\bm{55.3}$ & 16.5 & 1.7 & 0.1 \\
              & log-log & 1.2 & 31.3 & $\bm{53.5}$ & 12.1 & 1.7 & 0.1 \\
              & clog-log & 0.3 & 24.5 & $\bm{55.9}$ & 17.2 & 1.9 & 0.1 \\ \hdashline
    scenario 6 & PAVA & 0.2 & 2.6 & 26.3 & $\bm{49.4}$ & 19.6 & 1.9 \\
              & logit & 0.0 & 1.1 & 20.9 & $\bm{54.7}$ & 20.7 & 2.6 \\
              & log-log & 0.0 & 1.8 & 27.8 & $\bm{49.8}$ & 18.0 & 2.6 \\
              & clog-log & 0.0 & 1.0 & 20.0 & $\bm{54.1}$ & 21.7 & 3.2 \\ \hdashline
    scenario 7 & PAVA & 20.0 & $\bm{61.9}$ & 16.9 & 1.1 & 0.0 & 0.0 \\
              & logit & 4.0 & $\bm{75.4}$ & 19.9 & 0.6 & 0.0 & 0.0 \\
              & log-log & 6.6 & $\bm{76.6}$ & 16.1 & 0.6 & 0.0 & 0.0 \\
              & clog-log & 3.8 & $\bm{74.7}$ & 20.4 & 1.0 & 0.0 & 0.0 \\ \hdashline
    scenario 8 & PAVA & 3.1 & 29.5 & $\bm{52.7}$ & 13.3 & 1.3 & 0.0 \\ 
              & logit & 0.3 & 26.6 & $\bm{57.2}$ & 15.0 & 0.8 & 0.0 \\ 
              & log-log & 1.1 & 32.0 & $\bm{55.8}$ & 10.2 & 0.8 & 0.0 \\ 
              & clog-log & 0.3 & 25.1 & $\bm{57.9}$ & 15.8 & 0.8 & 0.0 \\ \hline
    \end{tabular}
\end{table}

\begin{table}[H]
    \centering
    \caption{MTD selection results for $p_1 = 0.7$ with $p_J = 0.21$. Prior distributions: logit: $\beta_0 \sim N(0.074, 1.713)$, $\beta_1 \sim N(-1.638, 1.087)$; log-log: $\beta_0 \sim N(0.410, 1.412)$, $\beta_1 \sim N(-2.330, 0.543)$; clog-log: $\beta_0 \sim N(-0.340, 1.532)$, $\beta_1 \sim N(-1.512, 0.500)$.} \label{tab:p1-4} 
    \begin{tabular}{cc cccccc} \hline
    \multicolumn{2}{c}{Dose levels} & 1 & 2 & 3 & 4 & 5 & 6 \\ 
    \multicolumn{2}{c}{Doses} & 10 & 20 & 30 & 45 & 60 & 80 \\ \hline 
    scenario 1 & PAVA & 0.9 & 9.2 & 28.1 & $\bm{33.8}$ & 24.6 & 3.4 \\ 
              & logit & 0.1 & 7.4 & 23.0 & $\bm{36.3}$ & 23.1 & 10.1 \\ 
              & log-log & 1.6 & 8.1 & 17.6 & $\bm{33.9}$ & 23.6 & 15.2 \\ 
              & clog-log & 0.2 & 7.8 & 19.3 & $\bm{35.8}$ & 22.4 & 14.5 \\ \hdashline
    scenario 2 & PAVA & 0.0 & 0.7 & 7.4 & 14.6 & 26.6 & $\bm{50.7}$ \\ 
              & logit & 0.0 & 0.0 & 4.9 & 13.9 & 20.1 & $\bm{61.1}$ \\ 
              & log-log & 0.0 & 0.0 & 4.5 & 13.0 & 18.5 & $\bm{64.0}$ \\ 
              & clog-log & 0.0 & 0.0 & 4.3 & 13.4 & 18.4 & $\bm{63.9}$ \\ \hdashline
    scenario 3 & PAVA & 0.0 & 0.2 & 6.9 & 27.4 & $\bm{56.7}$ & 8.8 \\ 
              & logit & 0.0 & 0.0 & 2.9 & 25.5 & $\bm{42.4}$ & 29.2 \\ 
              & log-log & 0.1 & 0.9 & 3.5 & 17.8 & $\bm{37.8}$ & 39.9 \\ 
              & clog-log & 0.0 & 0.3 & 3.1 & 20.7 & $\bm{36.6}$ & 39.3 \\ \hdashline
    scenario 4 & PAVA & 19.6 & $\bm{46.8}$ & 27.2 & 4.5 & 0.8 & 0.0 \\ 
              & logit & 20.0 & $\bm{49.2}$ & 22.7 & 6.2 & 0.8 & 0.0 \\ 
              & log-log & 32.2 & $\bm{36.6}$ & 20.6 & 8.1 & 1.3 & 0.1 \\ 
              & clog-log & 24.0 & $\bm{44.2}$ & 21.8 & 7.8 & 1.0 & 0.1 \\ \hdashline
    scenario 5 & PAVA & 3.1 & 29.2 & $\bm{51.0}$ & 14.1 & 2.4 & 0.1 \\
              & logit & 1.8 & 27.0 & $\bm{49.1}$ & 18.4 & 3.5 & 0.1 \\
              & log-log & 10.8 & 21.5 & $\bm{36.8}$ & 25.5 & 4.5 & 0.8 \\
              & clog-log & 5.1 & 24.7 & $\bm{40.9}$ & 24.3 & 4.3 & 0.6 \\ \hdashline
    scenario 6 & PAVA & 0.2 & 2.6 & 26.3 & $\bm{49.4}$ & 19.6 & 1.9 \\
              & logit & 0.0 & 1.6 & 18.7 & $\bm{48.3}$ & 25.0 & 6.4 \\
              & log-log & 1.4 & 4.0 & 13.3 & $\bm{41.3}$ & 29.8 & 10.2 \\
              & clog-log & 0.2 & 2.7 & 16.0 & $\bm{43.6}$ & 27.7 & 9.8 \\ \hdashline
    scenario 7 & PAVA & 20.0 & $\bm{61.9}$ & 16.9 & 1.1 & 0.0 & 0.0 \\
              & logit & 14.5 & $\bm{66.1}$ & 17.9 & 1.3 & 0.1 & 0.0 \\
              & log-log & 33.0 & $\bm{44.1}$ & 20.7 & 1.9 & 0.2 & 0.0 \\
              & clog-log & 23.3 & $\bm{53.5}$ & 21.3 & 1.7 & 0.1 & 0.0 \\ \hdashline
    scenario 8 & PAVA & 3.1 & 29.5 & $\bm{52.7}$ & 13.3 & 1.3 & 0.0 \\ 
              & logit & 1.8 & 27.8 & $\bm{50.4}$ & 18.1 & 1.8 & 0.0 \\ 
              & log-log & 11.5 & 22.4 & $\bm{36.8}$ & 25.6 & 3.5 & 0.1 \\ 
              & clog-log & 5.3 & 26.2 & $\bm{40.8}$ & 24.4 & 3.1 & 0.1 \\ \hline
    \end{tabular}
\end{table}

\begin{table}[H]
    \centering
    \caption{MTD selection results for $p_J = 0.08$ with $p_1 = 0.05$. Prior distributions: logit: $\beta_0 \sim N(-1.073, 1.128)$, $\beta_1 \sim N(0.687, 0.500)$; log-log: $\beta_0 \sim N(-0.196, 1.217)$, $\beta_1 \sim N(0.328, 0.569)$; clog-log: $\beta_0 \sim N(-1.255, 0.932)$, $\beta_1 \sim N(0.092, 0.642)$.} \label{tab:pJ-1} 
    \begin{tabular}{cc cccccc} \hline
    \multicolumn{2}{c}{Dose levels} & 1 & 2 & 3 & 4 & 5 & 6 \\ 
    \multicolumn{2}{c}{Doses} & 10 & 20 & 30 & 45 & 60 & 80 \\ \hline 
    scenario 1 & PAVA & 0.9 & 9.2 & 28.1 & $\bm{33.8}$ & 24.6 & 3.4 \\ 
              & logit & 0.0 & 7.5 & 31.2 & $\bm{41.2}$ & 18.8 & 1.3 \\ 
              & log-log & 0.0 & 9.5 & 32.9 & $\bm{40.4}$ & 16.2 & 1.0 \\ 
              & clog-log & 0.0 & 7.3 & 29.7 & $\bm{41.3}$ & 19.7 & 2.0 \\ \hdashline
    scenario 2 & PAVA & 0.0 & 0.7 & 7.4 & 14.6 & 26.6 & $\bm{50.7}$ \\ 
              & logit & 0.0 & 0.0 & 7.2 & 18.0 & 31.2 & $\bm{43.6}$ \\ 
              & log-log & 0.0 & 0.0 & 7.6 & 19.6 & 31.0 & $\bm{41.8}$ \\ 
              & clog-log & 0.0 & 0.0 & 6.9 & 16.9 & 29.4 & $\bm{46.8}$ \\ \hdashline
    scenario 3 & PAVA & 0.0 & 0.2 & 6.9 & 27.4 & $\bm{56.7}$ & 8.8 \\ 
              & logit & 0.0 & 0.0 & 4.8 & 37.4 & $\bm{52.3}$ & 5.5 \\ 
              & log-log & 0.0 & 0.0 & 5.7 & 36.5 & $\bm{52.9}$ & 4.9 \\ 
              & clog-log & 0.0 & 0.0 & 4.2 & 35.7 & $\bm{52.8}$ & 7.3 \\ \hdashline
    scenario 4 & PAVA & 19.6 & $\bm{46.8}$ & 27.2 & 4.5 & 0.8 & 0.0 \\ 
              & logit & 7.9 & $\bm{58.1}$ & 28.3 & 4.5 & 0.1 & 0.0 \\ 
              & log-log & 15.7 & $\bm{56.8}$ & 23.8 & 2.5 & 0.1 & 0.0 \\ 
              & clog-log & 7.2 & $\bm{57.0}$ & 29.8 & 4.8 & 0.1 & 0.0 \\ \hdashline
    scenario 5 & PAVA & 3.1 & 29.2 & $\bm{51.0}$ & 14.1 & 2.4 & 0.1 \\
              & logit & 0.2 & 26.1 & $\bm{58.8}$ & 14.0 & 0.8 & 0.0 \\
              & log-log & 1.2 & 32.9 & $\bm{53.3}$ & 11.7 & 0.8 & 0.0 \\
              & clog-log & 0.2 & 24.3 & $\bm{59.9}$ & 14.6 & 0.9 & 0.0 \\ \hdashline
    scenario 6 & PAVA & 0.2 & 2.6 & 26.3 & $\bm{49.4}$ & 19.6 & 1.9 \\
              & logit & 0.0 & 0.9 & 27.3 & $\bm{57.2}$ & 13.8 & 0.8 \\
              & log-log & 0.0 & 1.4 & 30.9 & $\bm{54.6}$ & 12.5 & 0.6 \\
              & clog-log & 0.0 & 0.9 & 25.2 & $\bm{56.9}$ & 16.0 & 1.0 \\ \hdashline
    scenario 7 & PAVA & 20.0 & $\bm{61.9}$ & 16.9 & 1.1 & 0.0 & 0.0 \\
              & logit & 1.7 & $\bm{75.6}$ & 22.3 & 0.3 & 0.0 & 0.0 \\
              & log-log & 6.7 & $\bm{76.4}$ & 16.5 & 0.3 & 0.0 & 0.0 \\
              & clog-log & 1.6 & $\bm{74.4}$ & 23.4 & 0.5 & 0.0 & 0.0 \\ \hdashline
    scenario 8 & PAVA & 3.1 & 29.5 & $\bm{52.7}$ & 13.3 & 1.3 & 0.0 \\ 
              & logit & 0.2 & 26.2 & $\bm{61.4}$ & 11.9 & 0.2 & 0.0 \\ 
              & log-log & 1.2 & 33.4 & $\bm{55.3}$ & 9.8 & 0.2 & 0.0 \\ 
              & clog-log & 0.2 & 24.4 & $\bm{62.2}$ & 12.8 & 0.3 & 0.0 \\ \hline
    \end{tabular}
\end{table}

\begin{table}[H]
    \centering
    \caption{MTD selection results for $p_J = 0.11$ with $p_1 = 0.05$. Prior distributions: logit: $\beta_0 \sim N(-0.939, 1.407)$, $\beta_1 \sim N(0.565, 0.805)$; log-log: $\beta_0 \sim N(-0.095, 0.906)$, $\beta_1 \sim N(0.085, 0.520)$; clog-log: $\beta_0 \sim N(-1.391, 0.939)$, $\beta_1 \sim N(0.310, 0.635)$.} \label{tab:pJ-2} 
    \begin{tabular}{cc cccccc} \hline
    \multicolumn{2}{c}{Dose levels} & 1 & 2 & 3 & 4 & 5 & 6 \\ 
    \multicolumn{2}{c}{Doses} & 10 & 20 & 30 & 45 & 60 & 80 \\ \hline 
    scenario 1 & PAVA & 0.9 & 9.2 & 28.1 & $\bm{33.8}$ & 24.6 & 3.4 \\ 
              & logit & 0.0 & 8.2 & 29.6 & $\bm{39.5}$ & 19.6 & 3.1 \\ 
              & log-log & 0.0 & 8.8 & 32.8 & $\bm{40.8}$ & 16.6 & 1.0 \\ 
              & clog-log & 0.0 & 7.2 & 29.2 & $\bm{40.0}$ & 20.5 & 3.1 \\ \hdashline
    scenario 2 & PAVA & 0.0 & 0.7 & 7.4 & 14.6 & 26.6 & $\bm{50.7}$ \\ 
              & logit & 0.0 & 0.0 & 7.7 & 15.6 & 27.1 & $\bm{49.6}$ \\ 
              & log-log & 0.0 & 0.0 & 7.4 & 19.2 & 31.0 & $\bm{42.4}$ \\ 
              & clog-log & 0.0 & 0.0 & 7.2 & 15.6 & 27.5 & $\bm{49.7}$ \\ \hdashline
    scenario 3 & PAVA & 0.0 & 0.2 & 6.9 & 27.4 & $\bm{56.7}$ & 8.8 \\ 
              & logit & 0.0 & 0.0 & 6.0 & 31.9 & $\bm{53.1}$ & 9.0 \\ 
              & log-log & 0.0 & 0.0 & 5.0 & 37.3 & $\bm{52.4}$ & 5.3 \\ 
              & clog-log & 0.0 & 0.0 & 5.0 & 32.3 & $\bm{53.4}$ & 9.3 \\ \hdashline
    scenario 4 & PAVA & 19.6 & $\bm{46.8}$ & 27.2 & 4.5 & 0.8 & 0.0 \\ 
              & logit & 11.3 & $\bm{56.2}$ & 26.8 & 4.5 & 0.1 & 0.0 \\ 
              & log-log & 13.5 & $\bm{57.9}$ & 25.1 & 2.3 & 0.1 & 0.0 \\ 
              & clog-log & 7.1 & $\bm{55.3}$ & 31.4 & 5.0 & 0.1 & 0.0 \\ \hdashline
    scenario 5 & PAVA & 3.1 & 29.2 & $\bm{51.0}$ & 14.1 & 2.4 & 0.1 \\
              & logit & 0.4 & 28.1 & $\bm{55.9}$ & 14.5 & 0.9 & 0.1 \\
              & log-log & 0.5 & 32.2 & $\bm{55.0}$ & 11.4 & 0.8 & 0.0 \\
              & clog-log & 0.1 & 23.5 & $\bm{59.4}$ & 15.8 & 1.0 & 0.1 \\ \hdashline
    scenario 6 & PAVA & 0.2 & 2.6 & 26.3 & $\bm{49.4}$ & 19.6 & 1.9 \\
              & logit & 0.0 & 1.4 & 26.2 & $\bm{54.8}$ & 16.0 & 1.6 \\
              & log-log & 0.0 & 1.0 & 30.0 & $\bm{55.6}$ & 13.0 & 0.4 \\
              & clog-log & 0.0 & 0.9 & 24.8 & $\bm{55.8}$ & 16.9 & 1.6 \\ \hdashline
    scenario 7 & PAVA & 20.0 & $\bm{61.9}$ & 16.9 & 1.1 & 0.0 & 0.0 \\
              & logit & 3.3 & $\bm{76.2}$ & 19.9 & 0.5 & 0.0 & 0.0 \\
              & log-log & 4.9 & $\bm{77.4}$ & 17.3 & 0.3 & 0.0 & 0.0 \\
              & clog-log & 1.6 & $\bm{72.7}$ & 24.9 & 0.7 & 0.0 & 0.0 \\ \hdashline
    scenario 8 & PAVA & 3.1 & 29.5 & $\bm{52.7}$ & 13.3 & 1.3 & 0.0 \\ 
              & logit & 0.4 & 28.3 & $\bm{58.2}$ & 12.6 & 0.4 & 0.0 \\ 
              & log-log & 0.5 & 32.6 & $\bm{57.0}$ & 9.7 & 0.1 & 0.0 \\ 
              & clog-log & 0.1 & 23.5 & $\bm{61.5}$ & 14.4 & 0.4 & 0.0 \\ \hline
    \end{tabular}
\end{table}

\begin{table}[H]
    \centering
    \caption{MTD selection results for $p_J = 0.3$ with $p_1 = 0.05$. Prior distributions: logit: $\beta_0 \sim N(-1.392, 1.358)$, $\beta_1 \sim N(0.322, 0.500)$; log-log: $\beta_0 \sim N(-0.229, 1.128)$, $\beta_1 \sim N(0.249, 0.500)$; clog-log: $\beta_0 \sim N(-1.655, 0.993)$, $\beta_1 \sim N(-0.140, 0.655)$.} \label{tab:pJ-3} 
    \begin{tabular}{cc cccccc} \hline
    \multicolumn{2}{c}{Dose levels} & 1 & 2 & 3 & 4 & 5 & 6 \\ 
    \multicolumn{2}{c}{Doses} & 10 & 20 & 30 & 45 & 60 & 80 \\ \hline 
    scenario 1 & PAVA & 0.9 & 9.2 & 28.1 & $\bm{33.8}$ & 24.6 & 3.4 \\ 
              & logit & 0.0 & 6.8 & 26.5 & $\bm{41.8}$ & 21.3 & 3.6 \\ 
              & log-log & 0.0 & 9.2 & 32.7 & $\bm{40.9}$ & 16.2 & 1.0 \\ 
              & clog-log & 0.0 & 6.2 & 25.2 & $\bm{39.9}$ & 23.3 & 5.4 \\ \hdashline
    scenario 2 & PAVA & 0.0 & 0.7 & 7.4 & 14.6 & 26.6 & $\bm{50.7}$ \\ 
              & logit & 0.0 & 0.0 & 5.8 & 16.0 & 26.3 & $\bm{51.9}$ \\ 
              & log-log & 0.0 & 0.0 & 7.5 & 19.7 & 31.0 & $\bm{41.8}$ \\ 
              & clog-log & 0.0 & 0.0 & 5.9 & 15.0 & 24.0 & $\bm{55.1}$ \\ \hdashline
    scenario 3 & PAVA & 0.0 & 0.2 & 6.9 & 27.4 & $\bm{56.7}$ & 8.8 \\ 
              & logit & 0.0 & 0.0 & 3.1 & 30.1 & $\bm{55.5}$ & 11.3 \\ 
              & log-log & 0.0 & 0.0 & 5.2 & 36.8 & $\bm{52.9}$ & 5.1 \\ 
              & clog-log & 0.0 & 0.0 & 3.1 & 27.8 & $\bm{54.1}$ & 15.0 \\ \hdashline
    scenario 4 & PAVA & 19.6 & $\bm{46.8}$ & 27.2 & 4.5 & 0.8 & 0.0 \\ 
              & logit & 8.6 & $\bm{56.0}$ & 29.1 & 5.1 & 0.1 & 0.0 \\ 
              & log-log & 15.4 & $\bm{57.0}$ & 23.9 & 2.5 & 0.1 & 0.0 \\ 
              & clog-log & 6.3 & $\bm{55.0}$ & 31.5 & 5.7 & 0.4 & 0.0 \\ \hdashline
    scenario 5 & PAVA & 3.1 & 29.2 & $\bm{51.0}$ & 14.1 & 2.4 & 0.1 \\
              & logit & 0.2 & 23.5 & $\bm{58.3}$ & 16.7 & 1.1 & 0.1 \\
              & log-log & 1.0 & 32.5 & $\bm{53.7}$ & 11.9 & 0.8 & 0.0 \\
              & clog-log & 0.1 & 21.0 & $\bm{57.0}$ & 19.6 & 2.1 & 0.1 \\ \hdashline
    scenario 6 & PAVA & 0.2 & 2.6 & 26.3 & $\bm{49.4}$ & 19.6 & 1.9 \\
              & logit & 0.0 & 0.9 & 20.8 & $\bm{57.8}$ & 18.9 & 1.6 \\
              & log-log & 0.0 & 1.0 & 30.2 & $\bm{55.4}$ & 13.0 & 0.4 \\
              & clog-log & 0.0 & 0.9 & 19.5 & $\bm{55.4}$ & 21.7 & 2.5 \\ \hdashline
    scenario 7 & PAVA & 20.0 & $\bm{61.9}$ & 16.9 & 1.1 & 0.0 & 0.0 \\
              & logit & 2.0 & $\bm{74.0}$ & 23.1 & 0.8 & 0.0 & 0.0 \\
              & log-log & 6.0 & $\bm{76.8}$ & 16.8 & 0.3 & 0.0 & 0.0 \\
              & clog-log & 1.5 & $\bm{71.9}$ & 25.3 & 1.2 & 0.0 & 0.0 \\ \hdashline
    scenario 8 & PAVA & 3.1 & 29.5 & $\bm{52.7}$ & 13.3 & 1.3 & 0.0 \\ 
              & logit & 0.2 & 23.7 & $\bm{60.5}$ & 14.8 & 0.7 & 0.0 \\ 
              & log-log & 1.0 & 33.0 & $\bm{55.7}$ & 10.0 & 0.2 & 0.0 \\ 
              & clog-log & 0.1 & 21.1 & $\bm{59.5}$ & 18.2 & 1.0 & 0.0 \\ \hline
    \end{tabular}
\end{table}

\newpage 

\subsubsection{Large variance prior for coefficient parameters}
We examined the impact of using large variance normal priors for the coefficient parameters $\beta_0$ and $\beta_1$ in the dose-response model \eqref{eq:drmodel}. 
This analysis contrasts with the main simulation results (Table 4 in the main manuscript), which employed carefully specified priors derived through our proposed approach in Section 3.2. 
Specifically, we replaced our default prior specifications, derived in Section 4.2, with normal distributions having large variances (variance $= 10$), while maintaining the same mean values. 
This represents approximately 10-fold increases in variance compared to the carefully calibrated specifications used in the main analysis. 
For example, for the logit link function, the large variance priors were:
$\beta_0 \sim N(-1.592, 10)$, $\beta_1 \sim N(0.412, 10)$ compared to the default specification $\beta_0 \sim N(-1.592, 1.371)$, $\beta_1 \sim N(0.412, 0.784)$. 
This large variance specification leads to U-shaped prior distributions for DLT probabilities at intermediate dose levels, where probability mass concentrates near the extremes (0 and 1) rather than moderate values typically expected in clinical practice. 

Table \ref{tab:flat_results} presents the MTD selection results under this prior specification across all eight scenarios. 
Comparing these results with Table 4 in the main manuscript reveals several important patterns. 
The large variance priors demonstrate markedly different performance depending on the location of the true MTD. 
In scenarios where the true MTD lies at boundary dose levels (e.g., Scenario 2 with MTD at the highest dose level 6), the large variance priors sometimes yield favorable results. 
For example, in Scenario 2, the logit model with large variance priors achieves 56.2\% correct MTD selection compared to 51.6\% with our default specification, representing a 4.6 percentage point improvement. 
This occurs because the U-shaped prior distribution, which concentrates probability mass at extreme values (near 0 and 1), aligns well with scenarios where the MTD is expected to be at the boundary. 

However, in all scenarios except Scenario 2, where the MTD is at the highest dose, the large variance priors consistently deteriorate performance. 
For example, in Scenario 1 (MTD at dose level 4), the proportion of correct MTD selections for the logit model drops from 39.5\% with default priors to 34.3\% with large variance priors, representing a 5.2 percentage point decrease. 
More substantial reductions exceed 10 percentage points in Scenarios 5 and 8, where the proportions drop from 57.9\% to 47.5\% and from 60.1\% to 49.4\%, respectively. 
Overall, across all seven scenarios with intermediate MTDs, the large variance priors result in an average decrease of 7.5 percentage points in MTD selection accuracy for the logit model, with similar deterioration patterns consistently observed across all three link functions (logit, log-log, and complementary log-log). 
This demonstrates that the adverse effects of inappropriate prior specification are not limited to a particular modeling choice but represent a fundamental issue with U-shaped prior distributions for intermediate dose levels. 
The deterioration occurs because the U-shaped prior distribution conflicts with the clinical expectation of moderate toxicity rates at intermediate dose levels, where the true MTD typically resides in Phase I oncology trials.

\begin{table}[H]
    \centering
    \caption{The proportion of MTD selections across 1,000 simulations in each scenario for the PAVA-based method and the proposed methods using the logit, log-log, and complementary log-log models. 
    The optimal MTD is highlighted in bold.} \label{tab:flat_results} 
    \begin{tabular}{cc cccccc} \hline
    \multicolumn{2}{c}{Dose levels} & 1 & 2 & 3 & 4 & 5 & 6 \\ 
    \multicolumn{2}{c}{Doses} & 10 & 20 & 30 & 45 & 60 & 80 \\ \hline 
    scenario 1 & PAVA & 0.9 & 9.2 & 28.1 & $\bm{33.8}$ & 24.6 & 3.4 \\ 
              & logit & 0.8 & 7.7 & 24.7 & $\bm{34.3}$ & 23.7 & 8.8 \\ 
              & log-log & 0.8 & 8.5 & 24.6 & $\bm{35.0}$ & 22.9 & 8.2 \\ 
              & clog-log & 0.8 & 7.5 & 24.8 & $\bm{33.0}$ & 25.3 & 8.6 \\ \hdashline
    scenario 2 & PAVA & 0.0 & 0.7 & 7.4 & 14.6 & 26.6 & $\bm{50.7}$ \\ 
              & logit & 0.0 & 0.4 & 6.9 & 14.0 & 22.5 & $\bm{56.2}$ \\ 
              & log-log & 0.0 & 0.5 & 6.9 & 14.1 & 23.0 & $\bm{55.5}$ \\ 
              & clog-log & 0.0 & 0.4 & 6.6 & 13.9 & 22.7 & $\bm{56.4}$ \\ \hdashline
    scenario 3 & PAVA & 0.0 & 0.2 & 6.9 & 27.4 & $\bm{56.7}$ & 8.8 \\ 
              & logit & 0.0 & 0.1 & 5.5 & 25.7 & $\bm{52.8}$ & 15.9 \\ 
              & log-log & 0.0 & 0.1 & 5.8 & 27.1 & $\bm{52.4}$ & 14.6 \\ 
              & clog-log & 0.0 & 0.1 & 5.1 & 24.3 & $\bm{54.0}$ & 16.5 \\ \hdashline
    scenario 4 & PAVA & 19.6 & $\bm{46.8}$ & 27.2 & 4.5 & 0.8 & 0.0 \\ 
              & logit & 18.3 & $\bm{47.6}$ & 25.2 & 6.9 & 0.9 & 0.0 \\ 
              & log-log & 21.8 & $\bm{46.5}$ & 24.0 & 5.7 & 0.9 & 0.0 \\ 
              & clog-log & 18.4 & $\bm{45.2}$ & 27.1 & 7.1 & 1.1 & 0.0 \\ \hdashline
    scenario 5 & PAVA & 3.1 & 29.2 & $\bm{51.0}$ & 14.1 & 2.4 & 0.1 \\
              & logit & 3.3 & 26.1 & $\bm{47.5}$ & 19.3 & 3.5 & 0.2 \\
              & log-log & 3.8 & 28.0 & $\bm{47.1}$ & 17.3 & 3.5 & 0.2 \\
              & clog-log & 3.3 & 25.3 & $\bm{47.1}$ & 20.6 & 3.4 & 0.2 \\ \hdashline
    scenario 6 & PAVA & 0.2 & 2.6 & 26.3 & $\bm{49.4}$ & 19.6 & 1.9 \\
              & logit & 0.2 & 2.0 & 21.2 & $\bm{48.4}$ & 23.2 & 5.0 \\
              & log-log & 0.2 & 2.4 & 22.2 & $\bm{49.1}$ & 21.4 & 4.7 \\
              & clog-log & 0.2 & 2.0 & 20.1 & $\bm{47.6}$ & 25.0 & 5.1 \\ \hdashline
    scenario 7 & PAVA & 20.0 & $\bm{61.9}$ & 16.9 & 1.1 & 0.0 & 0.0 \\
              & logit & 16.7 & $\bm{63.9}$ & 17.9 & 1.3 & 0.1 & 0.0 \\
              & log-log & 18.5 & $\bm{63.1}$ & 17.1 & 1.1 & 0.1 & 0.0 \\
              & clog-log & 16.0 & $\bm{62.6}$ & 19.9 & 1.3 & 0.1 & 0.0 \\ \hdashline
    scenario 8 & PAVA & 3.1 & 29.5 & $\bm{52.7}$ & 13.3 & 1.3 & 0.0 \\ 
              & logit & 3.3 & 26.4 & $\bm{49.4}$ & 18.7 & 2.1 & 0.0 \\ 
              & log-log & 3.9 & 28.4 & $\bm{48.5}$ & 17.3 & 1.8 & 0.0 \\ 
              & clog-log & 3.3 & 25.6 & $\bm{49.1}$ & 19.7 & 2.2 & 0.0 \\ \hline
    \end{tabular}
\end{table}

\bibliographystyle{apalike} 
\bibliography{references.bib}

\end{document}